\documentclass [11pt,a4paper]{article}
\usepackage{jcappub}

\usepackage{amssymb}
\usepackage{mathrsfs}

\newcommand{\doij}[5]{\href{http://dx.doi.org/#1}{{\cob #2 #3(#5)#4}}}
\newcommand{\arX}[1]{\href{http://arxiv.org/abs/#1}{{\ttfamily\com arXiv:#1}}}
\def\cob{\color{blue}}
\def\com{\color{magenta}}

\begin{document}
{\renewcommand{\thefootnote}{\fnsymbol{footnote}}

\title{Anomaly-free cosmological perturbations\\ in effective canonical quantum gravity}

\author[a]{Aurelien Barrau,}
\affiliation[a]{Laboratoire de Physique Subatomique et de Cosmologie,  Universit\'{e} Grenoble-Alpes, CNRS/IN2P3 53 av.\ des Martyrs, 38025 Grenoble cedex, France}
\emailAdd{barrau@in2p3.fr}

\author[b]{Martin Bojowald,}
\affiliation[b]{Institute for Gravitation and the Cosmos, The Pennsylvania State University, 104 Davey Lab, University Park, PA 16802, USA}
\emailAdd{bojowald@gravity.psu.edu}

\author[c]{Gianluca Calcagni,}
\affiliation[c]{Instituto de Estructura de la Materia, CSIC, Serrano 121, 28006 Madrid, Spain}
\emailAdd{calcagni@iem.cfmac.csic.es}

\author[d]{Julien Grain,}
\affiliation[d]{Institut d'Astrophysique Spatiale, CNRS/INSU, Universit\'e Paris-Sud 11, Orsay F-91405 France}
\emailAdd{julien.grain@ias.u-psud.fr}

\author[b,e]{Mikhail Kagan}
\affiliation[e]{Division of Science and Engineering, The Pennsylvania State University, Abington, 1600 Woodland Road, Abington, PA 19116, USA}
\emailAdd{mak411@psu.edu}

\setcounter{footnote}{0}

\abstract{
  This article lays out a complete framework for an effective theory of
  cosmological perturbations with corrections from canonical quantum
  gravity. Since several examples exist for quantum-gravity effects that
  change the structure of space-time, the classical perturbative treatment
  must be rethought carefully. The present discussion provides a unified
  picture of several previous works, together with new treatments of
  higher-order perturbations and the specification of initial states.}



\maketitle

\section{Introduction}

One of the most promising avenues towards tests of quantum gravity is the study
of the corrections it implies in cosmological perturbation equations. With
some luck, characteristic traces of high-energy space-time phenomena may be
left in the observable part of the cosmic microwave background. The derivation
of power spectra from different approaches to quantum gravity is therefore one
of the most active areas in the field. In this article, we consider the status
of developments in canonical quantum gravity, in particular loop quantum
gravity.

Canonical quantum gravity does not straightforwardly lead to cosmological
perturbation equations. Given a consistent quantum theory, one would have to
find a suitable class of semi-classical states, so that the expectation values
they define for basic operators (such as the spatial metric or its
inhomogeneous modes) are subject to equations of motion derived from the
Wheeler--DeWitt equation. The issue of finding semi-classical (or other
suitable) states in canonical quantum gravity is tricky --- for one, we are
unaware of any ground state that would be a good starting point for a
perturbative expansion. And if we assume that the problem can be solved,
consistent cosmological perturbation equations do not follow automatically,
for reasons closely related to the generally covariant nature and complicated
gauge content of the theory.

The classical theory has more equations than unknowns, and they appear as
different types. There are constraints that depend only on the modes and their
first time derivatives (or on configuration variables and their canonically
conjugate momenta), and there are equations of motion that determine
second-order time derivatives. The classical set of equations satisfies two
important consistency conditions: (i) The constraints, a spatial function $C$
and a vector field $\vec{D}$, are preserved by the equations of motion, so
that they need to be imposed only for initial values and then automatically
hold at all later times. This preservation of the constraints is the reason
why the set of equations for a smaller number of free functions is consistent
and allows the right number of non-trivial solutions. (ii) The constraints are
generators of a large set of gauge transformations which include space-time
Lie derivatives of phase-space functions, in particular their time
derivatives. Time derivatives of momentum variables imply the second-order
equations of motion; by being generated by gauge-inducing constraints, they
automatically preserve the constraints.

In canonical language, the equations of motion are generated by the
constraints, that is they follow from equations $\dot{f}=\{f,H[N,\vec{M}]\}$
in which the time derivative (along a direction with space-time components
related to the lapse function $N$ and the shift vector $\vec{M}$) of a mode
function $f$ is given by the Poisson bracket with a Hamiltonian
$H[N,\vec{M}]$. This Hamiltonian is a linear combination
$H[N,\vec{M}]=\int(NC+\vec{M}\cdot\vec{D})$ of the constraint functions $C$
and $\vec{D}$ and therefore must vanish when evaluated for solutions of the
field equations, for all $N$ and $\vec{M}$.  Property (i) is guaranteed if
$0=\dot{H}[N_1,\vec{M}_1]= \{H[N_1,\vec{M}_1], H[N,\vec{M}]\}$ for all $N_1$
and $\vec{M}_1$ and for all configurations for which $H[N_1,\vec{M}_1]=0$, in
which case the constraints are called first-class. The gauge transformations
mentioned in property (ii) take the form
$\delta_{(\epsilon_0,\vec{\epsilon})}f=
\{f,H[\epsilon_0,\vec{\epsilon}]\}$. For the specific choice of
$(\epsilon_0,\vec{\epsilon})=(N,\vec{M})$, one obtains time derivatives as
gauge transformations. The general gauge transformation with parameters
$(\epsilon_0,\vec{\epsilon})$ amounts to a Lie derivative along a space-time
vector field with components $(\epsilon_0/N, \vec{\epsilon}-
\epsilon_0\vec{M}/N)$. Property (ii) then has the following consequence: By
choosing different linear combinations, varying $N$ and $\vec{M}$ in the
Hamiltonian, one takes into account all possible time choices. In other words,
the theory is invariant under reparametrizations of time. It is therefore
possible to express the equations of motion in terms of only gauge-invariant
(i.e., coordinate independent) combinations of the modes. No gauge artifacts
couple to the physical degrees of freedom.

These consistency conditions are a specific realization of the general concept
of a first-class system of constraints, using the notation introduced by
Dirac. There is a set of constraints, in our case $H[N,\vec{M}]$, or
individually $C[N]:=\int NC=0$ and $\vec{D}[\vec{M}]:=\int
\vec{M}\cdot\vec{D}=0$, so that their Poisson brackets
$\{\vec{D}[\vec{M}_1],\vec{D}[\vec{M}_2]\}$, $\{C[N],\vec{D}[\vec{M}]\}$ and
$\{C[N_1],C[N_2]\}$ vanish whenever the constraints are satisfied. In the case
of gravity, the constraints obey the hypersurface-deformation algebra
\cite{DiracHamGR}
\begin{eqnarray}
 \{\vec{D}[\vec{M}_1],\vec{D}[\vec{M}_2]\} &=& -\vec{D}[{\cal
   L}_{\vec{M}_2}\vec{M}_1]\,,\label{DD}\\
\{C[N],\vec{D}[\vec{M}]\} &=& -C[{\cal L}_{\vec{M}}N]\label{HD}\,,\\
\{C[N_1],C[N_2]\} &=& \vec{D}[N_1\vec{\nabla}N_2-N_2\vec{\nabla}N_1]\,, \label{HH}
\end{eqnarray}
and are first-class (here, a metric is used to obtain the contravariant derivative $\vec \nabla$). The preservation of the constraints
$\dot{C}[N_1]=\{C[N_1],H[N,\vec{M}]\}$ and
$\dot{\vec{D}}[\vec{M}_1]=\{\vec{D}[\vec{M}_1],H[N,\vec{M}]\}$ then follows
directly, and the gauge-invariant modes $\phi$ are those that satisfy
$\{\phi,H[\epsilon_0,\vec{\epsilon}]\}=0$ (for all $\epsilon_0$ and
$\vec{\epsilon}$) whenever the constraints hold. The relations
(\ref{DD})--(\ref{HH}) have an interesting geometrical meaning
\cite{Regained}: They realize commutators of deformations of spacelike
hypersurfaces in space-time along vector fields $N\vec{n}+\vec{M}$ with unit
normals $\vec{n}$ to the hypersurfaces. Although this algebra does not refer
to the dynamics of the theory, it has a tight relation with it: Second-order
field equations for the metric which are covariant under symmetries obeying
the algebra (\ref{DD})--(\ref{HH}) must equal Einstein's equation (with an
unrestricted cosmological constant) \cite{Regained,LagrangianRegained}. This
result leaves only little room for quantum corrections. It is the canonical
analog of the familiar statement that a local covariant action can correct the
Einstein--Hilbert term only by higher-curvature contributions.

When quantum corrections are inserted in the equations of gravity, in
particular in cosmological perturbation equations, it is never clear whether
the delicate consistency conditions summarized in the first-class nature of
the constraint algebra remain intact. Especially background-independent
frameworks cannot directly rely on standard covariance arguments because their
notion of space-time, encoded in (\ref{DD})--(\ref{HH}), is supposed to emerge
in some way from solutions to their equations. The consistency of the
equations, however, must be ensured before they can be solved. If the approach
used as well as the methods employed to derive semi-classical equations are
covariant, the preserved symmetry implies consistent equations based on
first-class constraints. This statement could, for instance, apply to
effective equations derived from a path-integral quantization of gravity,
provided one uses the correct integration measure to make the theory
anomaly-free. (The latter condition is highly non-trivial.) In canonical
quantum gravity, however, consistency must be shown explicitly because the
different treatments of time and space derivatives in canonical quantizations
eliminate manifest covariance.

In our discussion so far we have assumed that the full field content of
  gravity is considered without fixing the space-time gauge. The issues we
  mention can formally be circumvented if one fixes the gauge before
  quantizing or before inserting quantum corrections in effective equations.
In many cases, gauge fixing before quantization can indeed be assumed to be
harmless, but the situation considered here is different. First, the
constraints we are dealing with are more complicated functions than, say, the
Gauss constraint of Yang--Mills theories. It is therefore more likely that the
constraints receive significant quantum corrections. If the constraints are
quantum corrected, the gauge transformations they generate are not of the
classical form. Gauge fixing before quantization is then inconsistent, because
one would fix the gauge according to transformations which subsequently will
be modified. Secondly, in the present case the dynamics is part of the gauge
system. A consistent theory must therefore quantize gauge transformations and
the dynamics at the same time; one cannot fix one part (the gauge) in order to
derive the second part (the dynamics) in an unrestricted way. Therefore, in
canonical quantum gravity one must consider the full space-time gauge algebra
without restrictions, or else one cannot be sure that the resulting theory is
consistent.

We should note that a less severe point of view, sometimes advocated in the
literature, is to regard gauge fixing prior to quantization as part of the
definition of a quantum theory which may turn out to be inequivalent to a
  non-gauge fixed quantum theory. In all cases where such a model can be
  checked to be self-consistent (as partially done in the hybrid cases
  mentioned in section\ \ref{hyb}), this attitude is legitimate. The resulting
  dynamics and physical predictions are, in general, quantitatively different
  from the theory quantized without fixing the gauge. Whenever available,
  however, the latter should be preferred because it implements the full
  system.

Another method that avoids dealing with the full system
  (\ref{DD})--(\ref{HH}) is reduced phase-space quantization, often combined
  with a technique called deparametrization. (Examples for cosmological
  perturbations have been provided in
  \cite{PertObsI,PertObsII,BKdustI,BKdustII}.) For the reduced phase space, one
  solves the constraints classically and computes all observables invariant
  under gauge transformations. The resulting phase space is then to be
  quantized, which in general can be very complicated but may be possible in
  reduced or perturbative models. To facilitate the construction of
  gauge-invariant observables in a relational form, deparametrization selects
  a suitable phase-space degree of freedom to play the role of time. Formally,
  such a procedure leads to consistent equations, but they correspond to a
  quantized dynamics for classical observables rather than a complete
  quantization of the original gauge system. Moreover, for meaningful results
  one should check that the choice of internal time does not affect
  predictions, a form of covariance problem which is rarely analyzed in this
  context (see \cite{ReducedKasner}).

The preferred method in our view is to quantize the constraints without
  classical specifications of gauge or observables.  Canonical quantum
gravity then provides operator versions of the classical constraints $C[N]$
and $\vec{D}[\vec{M}]$. Consistency is formulated just as for the classical
constraints, except that commutators are used instead of Poisson brackets to
define a first-class system. Making sure that such an operator algebra is
first class is even more difficult than showing this property for a classical
Poisson-bracket algebra because the result is usually sensitive to the
particular factor ordering or possible regularization schemes used to define
the operators. (A closed algebra may even be in conflict with Hermitian
constraint operators \cite{Komar,NonHerm}.) It is therefore useful to combine
the derivation of a consistent algebra with the one of effective or
semi-classical equations. Such an approach cannot prove that any particular
theory of quantum gravity is consistent, due to the more-involved nature of
operator algebras. But it can show what kind of geometrical or physical
implications quantized constraints can have, including the potential for
observational tests in cosmology.

With such an effective viewpoint, one (i) takes the corrections
suggested by operator definitions in some approach to quantum gravity,
(ii) parametrizes them so as to allow for sufficient freedom to
encompass the ambiguities and unknowns in quantum operators, (iii)
inserts them in the classical constraints and (iv) computes their
algebra under Poisson brackets. In most cases, the algebra will no
longer be first class, but careful choices of the functions and
parameters used to specify the modifications may respect this
important feature.  If no choice of parameters gives rise to a
first-class algebra, the quantum effect described by them is likely to
be inconsistent. If there are consistent choices (which, in general,
are not unique; see section\ \ref{fref}), it is possible for the quantum
effect studied to be part of an anomaly-free theory of quantum
gravity, and one obtains a consistent model whose equations can be
analyzed for further implications. In particular, one can then derive
a complete set of cosmological perturbation equations and observables,
incorporating quantum corrections. (At this stage, consistency being
assured, one may use gauge fixing of the quantum-corrected constrained
system to simplify solving the equations of motion.)

The prescription just sketched has been defined and evaluated in detail in
\cite{ConstraintAlgebra}, with applications to a certain type of corrections
suggested by loop quantum gravity. In these cases, the algebra of constraints
remained first class, but it was deformed: Its structure functions in
(\ref{HH}), not just the constraints themselves, were affected by quantum
corrections. Two developments that have happened in the meantime motivate us
to take a renewed look: First, canonical effective theories and especially the
origin of higher time derivatives in them has been better
understood. Secondly, applied to a different type of modifications in loop
quantum gravity \cite{JR,ScalarHol}, the surprising (and in some eyes,
shocking) possibility of signature change \cite{Action} at high density was
found. As a consequence, the framework has been subjected to enhanced scrutiny
and criticism. This article presents a unified setting, combining insights
found in several papers since \cite{ConstraintAlgebra}, fleshing out some
details that may so far have been mentioned only rather implicitly, and
contrasting the approach with others that avoid dealing with the constraint
algebra and instead use gauge fixing or deparametrization before the theory is
quantized or modified.

We end this introduction with a summary of our main results. This article
  includes several passages of review in order to make especially the
  conceptual part of our discussions self-contained. A clear-cut separation
  between review material and novel statements would be rather artificial,
  since both are interwoven and the former is systematically utilized to
  introduce the latter. For the reader's convenience, however, we state here
  the main results in a nutshell:
\begin{enumerate}
\item The role of quantum moments in the effective dynamics is clarified and
  contrasted with that of holonomy modifications. The main difference is that
  higher-order time derivatives arise only by the first type of corrections
  (section \ref{s:Corr}). Section \ref{s:Degrees} contains a brief summary of
  effective methods, with special emphasis on the anomaly problem and the
  generality of this type of effective constructions. The techniques of
  relevance for our considerations do not make use of further approximations,
  such as derivative expansions, which would restrict the availability of
  effective descriptions.
\item By performing a detailed comparison of the most recent results on
  anomaly-free realizations of spherically symmetric models of loop quantum
  gravity as well as cosmological perturbations, we arrive at a simplified and
  more uniform picture of parametrizations of quantum corrections
  (section \ref{s:Examples}, especially sections \ref{323} and \ref{fref}).
\item 
The definition and possibility of signature change, which is an important ingredient for understanding the early universe, is discussed at length in section \ref{s:Sig}. In particular, we show that
  signature change, if it occurs, is not a consequence of perturbative inhomogeneity but
  rather of holonomy modifications used crucially in the background dynamics
  of loop quantum cosmology. We also contrast signature change with the more
  common phenomenon of instability.
\item Details are provided for systematic extensions of perturbation schemes
  to higher orders, paying due attention to degrees of freedom and symplectic
  structures (section \ref{s:Pert}). The role of gauge invariance and different
  applications of gauge fixing or deparametrization are discussed across
  various approaches to cosmological perturbations (section \ref{s:Add}).
\item When computing the inflationary spectra, the choice of vacuum is simply
  a choice of initial conditions consistent with the canonical effective
  equations. It is not a ``re-quantization'' of the theory, as sometimes
  hinted at in the literature. Our discussion of this question clarifies this
  issue and has the additional advantage of generalizing the specification of
  vacuum initial conditions to non-standard space-time structures as
  encountered in models of loop quantum gravity (section \ref{s:InflVac}).
\end{enumerate}

\section{Canonical quantum gravity}

Canonically, general relativity is described by the phase space
$(h_{ab},p^{ab})$ of spatial metrics $h_{ab}$ and momenta $p^{ab}$ related to
extrinsic curvature. The theory can then be parametrized conveniently using
the ADM decomposition \cite{ADM}. A space-time solution is
\begin{equation} \label{ds}
 {\rm d}s^2= -N^2{\rm d}t^2+ h_{ab} ({\rm d}x^a+M^a{\rm d}t) ({\rm
   d}x^b+M^b{\rm d}t)
\end{equation}
where $h_{ab}$ changes between $t={\rm const}$ slices by ${\cal L}_t
h_{ab}=\{h_{ab}, H[N,\vec{M}]\}$ together with ${\cal
  L}_tp^{ab}=\{p^{ab},H[N,\vec{M}]\}$. Initial values for $h_{ab}$ and
$p^{ab}$ must obey the constraints $H[N,\vec{M}]=0$. Time derivatives are
represented as Lie derivatives along the time-evolution vector field
$t^a=Nn^a+M^a$ with the unit normal $n^a$. Since all equations are invariant
under the transformations generated by $H[\epsilon_0,\vec{\epsilon}]$ and
(\ref{DD})--(\ref{HH}) imply that gauge transformations of fields obeying the
constraints are space-time Lie derivatives, (\ref{ds}) is invariant under
changes of coordinates.

Alternatively, one may use different variables that describe the same phase
space after a canonical transformation (and perhaps an enlargement by further
gauge degrees of freedom), such as Schwinger's \cite{Schwinger}
$(K_a^i,E^b_j)$ with the densitized triad $E^b_j=|\det (e_c^k)| e^b_j$ and the
variant $K_a^i=e^{ib}K_{ab}$ of extrinsic curvature, or Ashtekar--Barbero
variables \cite{AshVar,AshVarReell} $(A_a^i,E^b_j)$ with the connection
$A_a^i=\Gamma_a^i+\gamma K_a^i$, using the spin connection $\Gamma_a^i$ and
the real-valued Barbero--Immirzi parameter $\gamma$
\cite{AshVarReell,Immirzi}. The latter are used in loop quantum gravity
\cite{Rov,ThomasRev,ALRev}, which is based on a representation of holonomies
of $A_a^i$. For further details on canonical gravity, see \cite{CUP}.

\subsection{Classical theory}

In ADM variables, we have the relation 
\begin{equation}
 p^{ab}(x) =  \frac{\sqrt{\det h}}{16\pi G} (K^{ab}-K^c_ch^{ab})
\end{equation}
between the momentum and the extrinsic curvature $K_{ab}$. The variables are
subject to the Hamiltonian constraint
\begin{equation}
 C_{\rm grav}[N]= \int{\rm d}^3x\, N \left\{\frac{16\pi G}{\sqrt{\det
     h}} \left[p_{ab}p^{ab}- \frac{1}{2}(p^a_a)^2\right]-
 \frac{\sqrt{\det h}}{16\pi G}\, {}^{(3)}R\right\},
\end{equation}
with the spatial Ricci scalar ${}^{(3)}R$ computed from $h_{ab}$, and the
diffeomorphism constraint
\begin{equation}
\vec{D}_{\rm grav}[\vec{M}]= 2\int{\rm d}^3x\, M^a \sqrt{\det h} D^b
 \left[(\det h)^{-1/2} p_{ab}\right]=2\int{\rm d}^3x\, M^a D_bp_a^b
\end{equation}
where $D_a$ is the spatial covariant derivative. The full constraints are
obtained by adding the matter Hamiltonian to get $C[N]=C_{\rm grav}[N]+H_{\rm
  matter}[N]$, and an energy-flux contribution from matter to get
$\vec{D}[\vec{M}]= \vec{D}_{\rm grav}[\vec{M}]+\vec{D}_{\rm
  matter}[\vec{M}]$. The matter contribution to $\vec{D}[\vec{M}]$ is usually
of a simple form in which momenta are multiplied with spatial derivatives of
the configuration variables, such as $\int{\rm d}^3x\, M^a p_{\phi}D_a\phi$
for a scalar field. (For variables with density weight one, the derivative
will appear on the momentum.)

For Ashtekar--Barbero variables, we have the gravitational contribution to the
diffeomorphism constraint given by
\begin{equation}
 \vec{D}_{\rm grav}[\vec{M}]= \int{\rm d}^3x\, M^aP^b_j\left[
    F_{ab}^j+(1+\gamma^2)\epsilon^j{}_{kl}K_a^kK_b^l\right],
\end{equation}
with the Yang--Mills curvature $F_{ab}^i=2\partial_{[a}A^i_{b]}+ \epsilon^{ijk}
A_a^jA_b^k$ of $A_a^i$, and the contribution
\begin{equation} \label{CA}
 C_{\rm grav}[N]=4\pi G\gamma^2\int{\rm d}^3x\, N \frac{P^a_iP^b_j}{\sqrt{\det h}}
 \epsilon^{ij}{}_k \left[F_{ab}^k-(1+\gamma^2) \epsilon^k{}_{mn}
   K_a^mK_b^n\right]
\end{equation}
to the Hamiltonian constraint.  The momentum of $A_a^i$ is $P^a_i:=
E^a_i/(8\pi\gamma G)$, so that $\{A_a^i(x), P_j^b(y) \} = \delta_j^i
  \delta_a^b \delta (x,y)$.

Stress-energy components can be computed from the matter contributions to the
Hamiltonian and diffeomorphism constraint. The most immediate variables
are those measured by a so-called Euclidean observer, one that moves along
geodesics with 4-velocity given by the unit normals to spacelike hypersurfaces
$t={\rm const}$ in (\ref{ds}): the energy density, energy current and spatial
stress are (see, e.g., \cite{CUP,Energy})
\begin{equation} \label{rhoJS}
\rho_{\rm E}=\frac{1}{\sqrt{\det h}} \frac{\delta C_{\rm matter}[N]}{\delta
  N}, \quad J^{\rm E}_a= \frac{1}{\sqrt{\det h}} \frac{\delta \vec{D}_{\rm
    matter}[M^a]}{\delta 
   M^a},\quad S^{\rm E}_{ab}= -\frac{2}{N\sqrt{\det h}} 
\frac{\delta C_{\rm matter}[N]}{\delta h^{ab}}\,.
\end{equation}
The matter energy density measured by a generic
observer with 4-velocity ${\bf u}=(u^0,\vec{u})$ is then
\begin{equation}
\rho=(u^0)^2\rho_{\rm E}+2u^0u^aJ_a^{\rm E}+ u^au^bS_{ab}^{\rm E}\,.
\end{equation}

The expression for spatial stress $S^{\rm E}_{ab}$ contains a functional
derivative with respect to the spatial metric. In order to make a connection
with the loop cosmological perturbation theory, it is more convenient to
rewrite it in terms of the (densitized) triad. We start with the definition
\begin{equation}\label{def_h}
h^{ab}=\frac{E^a_i E^b_i}{|\det E|}\,,
\end{equation}
where $\det h = |\det E|$, and a useful auxiliary relation
\[
\delta(\det E) = \det E \, \underline E_a^i \,\delta E^a_i
\]
in terms of the inverse triad, given by
\begin{equation}
\underline E_a^i = \frac{\epsilon_{abc}\epsilon^{ijk}E^b_j
E^c_k}{3!\det E}\,.
\end{equation}
Using the above, the variation of the spatial metric can be rewritten as
\begin{equation}\label{delta_h}
\delta h^{ab}=\frac{1}{|\det E|} \left(E^a_i\delta E^b_i+E^b_i\delta
E^a_i-E^a_i E^b_i \underline E_c^k \delta E^c_k\right)\,.
\end{equation}

According to (\ref{rhoJS}), the variation of the matter constraint associated
with the variation of the spatial metric is
\begin{equation}\label{delta tildeC_def}
\delta C_{\rm matter}[N]=-\frac{1}{2}\int{\rm d}^3 x\,
N\sqrt{\det h} \, S^{\rm E}_{ab}\delta h^{ab}\,.
\end{equation}
Substituting (\ref{delta_h}) in the last expression and using the symmetry
of the spatial stress tensor, $S^{\rm E}_{ab}=S^{\rm E}_{ba}$, we obtain
\begin{equation}
\delta C_{\rm matter}[N]=-\frac{1}{2}\int{\rm d}^3 x\, \frac{N}{\sqrt{|\det
    E|}} S^{\rm E}_{ab} 
\left(2 E^a_i\delta E^b_i-E^a_i E^b_i \underline E_c^k \delta
E^c_k\right)\,.
\end{equation}
Taking a functional derivative with respect to $E_k^c$ yields
\begin{equation}\label{deltaC_deltaE}
\frac{\delta C_{\rm matter}[N]}{\delta E_k^c}=-\frac{N}{2\sqrt{|\det E|}}
\left(2S^{\rm E}_{ac} 
E^a_k-S^{\rm E}_{ab} E^a_i E^b_i \underline E_c^k \right)\,.
\end{equation}
The last term is proportional to the trace of the spatial stress tensor,
$S^{\rm E}_{ab} h^{ab}$, or the pressure
\begin{equation}\label{pressure_def}
P_{\rm E}= \frac{1}{3}S^{\rm E}_{ab} h^{ab} = \frac{2E^c_k}{3N\sqrt{|\det E|}}
\frac{\delta C_{\rm matter}[N]}{\delta E_k^c}\,.
\end{equation}
Substituting the trace back into (\ref{deltaC_deltaE}),
contracting with $E_k^b$, and solving for $S^{\rm E}_{ab}$, we obtain the
spatial stress tensor
\begin{eqnarray}\label{deltaC_deltaH}
S^{\rm E}_{ab} = -\frac{1}{N\sqrt{|\det E|}} h_{ac}\frac{\delta C_{\rm
    matter}[N]}{\delta E_i^d}\left(\delta_b^d   E_i^c-\delta_b^c E_i^d\right) \,,
\end{eqnarray}
where $h_{ab}$ is the inverse of (\ref{def_h}) as a function of the densitized
triad.

As shown in \cite{Energy}, the local conservation law of the stress-energy
tensor, which can canonically be derived from the algebra
(\ref{DD})--(\ref{HH}), no longer holds when the algebra is deformed. However,
a consistent form of continuity equations for the components (\ref{rhoJS})
remains intact. Specifically, the Poisson braket of (\ref{rhoJS}) with the
constraints should vanish weakly. If one does not have information on the
off-shell algebra, no affirmative statement about energy conservation can be
made.

\subsection{Quantization}

In order to quantize the theory canonically, one turns the phase-space
variables into basic operators acting on wave functions of a certain kind. In
terms of $(h_{ab},p^{ab})$, one usually employs formal constructions in which
``$\hat{h}_{ab}(x)$'' is a multiplication operator on a functional
$\psi[h_{ab}]$ and ``$\hat{p}^{ab}(x)$'' a functional derivative. Difficulties
in making sense of such tensorial operators of local fields can be overcome
with a loop quantization \cite{LoopRep}. As basic objects one introduces
holonomies $h_e(A)={\cal P}\exp(\int_e {\rm d}\lambda\,\tau_iA_a^i\dot{e}^a)$
(with Pauli matrices $2i\tau_i$) assigned to curves $e$ in space, and fluxes
$F^{(f)}_S(E)=\int_S{\rm d}^2y\,E^a_if^i$ assigned to surfaces $S$ in space
(and with su(2)-valued smearing functions $f^i$). Holonomies and fluxes, when
known for all curves and surfaces, allow one to reconstruct the local fields
$A_a^i$ and $E^b_j$. Instead of being space-time tensors, they take values in
the group SU(2) or its Lie algebra, and therefore have more manageable
transformation properties. Moreover, they are integrated or smeared in such a
way that a well-defined and non-distributional algebra under Poisson brackets
results (free of $\delta$-functions). Its representation on functions of
holonomies is the quantum representation used in loop quantum gravity.

It is now $h_e(A)$ that serves as a multiplication operator, for every curve
$e$ in space. Fluxes, quantizing the densitized triad,  are related to
derivative operators, taking the form of angular-momentum operators on
SU(2). They have discrete spectra, indicating that space acquires a discrete
geometric structure \cite{AreaVol,Area}.

The question of what geometric structure space-{\em time} has in this theory
is more complicated, and not completely solved. At this stage, the crucial
issues to be faced by canonical quantizations of gravity enter. Space-time
structure is determined by properties of the constraints $C$ and $\vec{D}$
that provide consistent evolution equations for the metric (or densitized
triad) and its modes, especially by the algebra (\ref{DD})--(\ref{HH}) or a
quantum version of it. At the same time, they generate gauge transformations
which classically take the form of coordinate changes in space-time. These
constraints must be turned into operators, but they depend on the connection
while only holonomies are represented in the Hilbert space of loop quantum
gravity. For quantized constraints, especially $C$, one therefore modifies
(or, using a common euphemism, ``regularizes'') the classical expression
before it can be represented as an operator \cite{RS:Ham,QSDI}. One uses
combinations of holonomies so that for short curves $e$ (or small curvature
components that enter the connection) they agree with the classical functional
of connection components. The precise classical expression is obtained only in
a limit, so that for any extended curve there are quantum corrections. They
can be understood as a power series
\begin{equation} \label{hol}
 h_e(A)= 1+\epsilon \tau_iA_a^i\dot{e}^a+\cdots\,,
\end{equation}
where $\epsilon$ is a small parameter depending on the coordinate length of
the curve $e$. In addition to higher powers of the connection, there are
higher spatial derivatives when the integration over $e$ is suitably
expanded. (The treatment of higher spatial derivatives has several subtleties;
see \cite{HigherSpatial}.)

It turns out that there is a second type of correction in loop quantum
gravity. The classical constraint $C$ depends on an inverse of the densitized
triad, or on its inverse determinant. Since the densitized triad is quantized
by flux operators with zero in their discrete spectra, there is no
densely-defined inverse operator. Nevertheless, one can quantize the inverse
in a more indirect way, making use of the classical identity \cite{QSDI}
\begin{equation}
 \left\{A_a^i,\int{\rm d}^3x\, \sqrt{|\det E^b_j|}\right\}\propto
 \epsilon^{ijk}\epsilon_{abc} \frac{E^b_jE^c_k}{\sqrt{|\det E^d_l|}}\,.
\end{equation}
On the left-hand side, no inverse of the densitized triad is required, and the
connection can be expressed in terms of holonomies. Once the Poisson bracket
is replaced by a commutator divided by $i\hbar$, a well-defined quantization
of the right-hand side is obtained. This commutator reproduces the right-hand side in the
classical limit for large flux eigenvalues, but shows characteristic
inverse-triad corrections \cite{InvScale,LoopMuk} for small values.

Holonomy modifications and inverse-triad corrections change the constraint $C$
and the dynamical equations of motion it generates. A more subtle effect is a
possible change in space-time structure because $C$ generates not only
equations of motion but also gauge transformations under which observables are
invariant. One ends up with a consistent set of equations and observables if the
algebra of constraints obtained after all modifications have been inserted
remains first class. Ideally, one would like to make sure that even the
underlying constraint operators obey a first-class commutator algebra, but
existing results remain incomplete. General arguments \cite{AnoFree} only
indicate that the commutators obey a version of (\ref{HH}) if the constraint
$\vec{D}$ is assumed to be solved. The right-hand side of (\ref{HH})
then vanishes, but one does not see the more-specific form of the actual
commutator for $\vec{D}\not=0$. These arguments therefore do not show complete
consistency \cite{LM:Vertsm,Consist,NPZRev}, nor do they elucidate the
possible form of quantum space-time structure in the theory.

There are more-detailed calculations that have shown a consistent version of
(\ref{HH}) at the operator level, so far in ($2+1$)-dimensional models
\cite{ThreeDeform,TwoPlusOneDef,TwoPlusOneDef2,AnoFreeWeak} and in
  spherically symmetric models \cite{SphSymmOp}. These results are
promising, but it remains unclear how easy it will be to extend them to the
full theory. Moreover, for phenomenological or effective equations one would
have to find suitable semi-classical states, so that a consistent operator
algebra, if it exists, could be translated into a consistent version of
quantum-corrected cosmological perturbation equations. Both steps seem very
difficult at present, and therefore it is important to test possible
consistent versions and their implications with effective methods such as
those described here, computing the algebra of modified classical constraints.

Also for modified classical constraints, several consistent versions have been
found. The set of all these calculations (operator and effective ones)
shows mutual consistency: In all cases, the classical algebra is deformed by
quantum corrections in the same way, changing (\ref{HH}) to 
\begin{equation} \label{HHbeta}
 \{C[N_1],C[N_2]\} = \vec{D}[\beta (N_1\vec{\nabla}N_2-N_2\vec{\nabla}N_1)]
\end{equation}
with a phase-space function $\beta$, while the other Poisson brackets
involving $\vec{D}$ remain unmodified. The form of $\beta$ depends only on
what quantum correction is implemented but not on the models or methods used
to derive the algebra. (See section \ref{s:Examples} for more details.) No
generic undeformed consistent version has been found, so that quantum
corrections in the constraint algebra, especially in (\ref{HH}), appear to be
a universal feature of models of loop quantum gravity.

\subsection{Quantum corrections}
\label{s:Corr}

Loop quantum gravity gives rise to holonomy modifications and inverse-triad
corrections, and other canonical approaches provide less clearly-defined
quantum terms resulting for instance from factor ordering or the
regularization of functional derivatives. Since loop modifications are
closely related to the basic representation and therefore very characteristic
of the theory, their detailed study is of considerable interest in order to
see if the theory can be falsified.

These loop corrections modify the classical constraints but do not introduce
new degrees of freedom: One just quantizes the classical phase space by
turning a preferred set of variables into operators. The form of the latter
then requires certain adaptations of the constraints that generate the
dynamics, and quantum corrections result. If momenta are related to first time
derivatives, upon using equations of motion, non-standard kinetic terms may be
obtained with higher powers of the first derivative (but no higher time
derivatives). Nowhere in this process do new degrees of freedom or
modifications of the phase space structures arise.

\subsubsection{Degrees of freedom and effective phase space}
\label{s:Degrees}

The preceding statement may come as a surprise because holonomy corrections
may look like (and are often taken\footnote{See, for instance, the
  statements in \cite{ActionRhoSquared}, expanded in \cite{ActionOrder}. While
  these papers provide candidates for higher-curvature actions that could
  capture some implications of holonomy modifications, they do not deal
  directly with holonomy terms or higher orders in a connection expansion such
  as (\ref{hol}). Instead, they take as their starting point a modified
  Friedmann equation of the form $H^2=(8\pi G/3) \rho(1-\rho/\rho_0)$ with
  some constant $\rho_0$, which in a restricted set of homogeneous and
  isotropic models has been shown to be equivalent to an effective Friedmann
  equation in the presence of a holonomy modification of the left-hand side
  $H^2$ \cite{AmbigConstr,APSII,BouncePert}. The $\rho^2$-correction can be
  obtained from an isotropic reduction of a suitable $F(R)$ action and could
  be taken as a higher-curvature theory corresponding to the isotropic
  holonomy-modified model. However, (i) the authors of
  \cite{ActionRhoSquared,ActionOrder} assume (but do not show) that
  inhomogeneous space-time can still be described by the classical notion of
  covariance and (ii) the relation between holonomy modifications and the
  simple $\rho^2$-correction breaks down when one goes beyond the simplest
  isotropic models. Therefore, \cite{ActionRhoSquared,ActionOrder} is not
  directly usable to understand the fundamental nature of space-time corresponding to
  holonomy-modified equations. A recent result shows that a careful
  consideration of isotropic models with modifications suggested by loop
  quantum gravity in fact indicates the need for modified space-time
  structures: The various versions of modified Friedmann equations suggested
  so far in loop quantum cosmology cannot be consistent with classical
  coordinate transformations \cite{MetricRestrict}.}  as examples of) familiar
higher-curvature corrections generically expected for quantum gravity
\cite{BurgessLivRev,EffectiveGR}. However, as already suggested by the origin
of holonomy modifications, they are not of higher-curvature type: Holonomy
corrections arise from considerations of {\em spatial} quantum geometry, not
from possible forms of {\em space-time} covariant actions. One would, of
course, expect quantum corrections to respect some kind of space-time
covariance for the theory to be consistent, but in canonical approaches this
is something to be shown, not to be assumed. Before covariance structures of
the modified constraints are analyzed (by studying the first-class nature of
their Poisson brackets), one cannot be sure that holonomy corrections are of
standard higher-curvature type.

Nevertheless, holonomy modifications provide corrections that depend on
certain measures of curvature. The connection is a linear combination of the
spin connection (related to intrinsic curvature of space) and extrinsic
curvature. Furthermore, one may, 
for dimensional reasons,
assume length parameters for curves along which holonomies are integrated to
be close to the Planck length (although this is not guaranteed). We then
assume that the spatial structure of an effective theory is the same as the
classical one. This assumption may, of course, be questioned, but it is weaker
than assuming the classical space-time structure. And it is supported by the
fact that the constraint $\vec{D}$ (or the finite transformations it
generates) does not receive holonomy modifications or inverse-triad
corrections in loop quantum gravity \cite{ALMMT}, while $C$ does. (A
modification of the theory may lead to corrections of $\vec{D}$
\cite{DiffeoOp}, but this possibility has not been fully analyzed.) With this
assumption, one would conclude that a combination of holonomies used to
``regulate'' the constraint $C$ has an expansion of the form $\sqrt{\det
  h}\sum_{n=0}^{\infty} \ell_{\rm P}^{2n}(\alpha_n \:{}^{(3)}\!R_n+ \beta_n
K_n)$ with the spatial metric $h_{ab}$ and spatial curvature invariants
${}^{(3)}\!R_n$ (depending on spatial derivatives of $h_{ab}$) as well as
spatial scalars $K_n$ formed from extrinsic curvature (that is, containing
time derivatives of $h_{ab}$ on shell), both of order $n$. (For $n=0$,
${}^{(3)}\!R_0$ is the 3-dimensional Ricci scalar.)  Holonomy modifications
therefore provide spatial higher-curvature terms, but not necessarily
space-time ones.

The previous statement is related to the question of degrees of
freedom. Higher-curvature terms of a space-time theory imply new degrees of
freedom because they always involve higher derivatives in time. (In an
initial-value formulation, not just the fields and their first time
derivatives must be chosen initially, but higher derivatives as well; this is
the so-called Ostrogradski problem.) Such new degrees of freedom would have to
modify the phase-space structure by an extension of the classical phase
space. If only higher spatial derivatives are present, however, as in expanded
holonomies, no new degrees of freedom arise, and the phase space does not
change.\footnote{In some theories such as Ho\v{r}ava--Lifshitz gravity
  \cite{Horava}, there are higher spatial but not higher time derivatives,
  which nevertheless give rise to new degrees of freedom. However, these
  degrees of freedom arise from a restriction of space-time diffeomorphisms to
  foliation-preserving diffeomorphisms, or from extra terms in the action with
  free auxiliary functions on top of the metric degrees of freedom. No such
  auxiliary functions are suggested by holonomy modifications.}  Holonomy
modifications merely amount to a redefinition of the dynamics and perhaps of
the space-time structure via a deformed constraint algebra. They do not modify
the phase space of an effective theory.

Like any interacting quantum theory, loop quantum gravity should generically
give rise to higher time derivatives and therefore new degrees of freedom in
an effective description. Indeed it does, but these corrections are not the
same as holonomy modifications. At a canonical level, higher-time derivatives
in an effective formulation of some quantum theory arise from coupling terms
between expectation values of basic operators with fluctuations and higher
moments \cite{EffAc,Karpacz,HigherTime}.  Appendix \ref{App:Moments} provides
a brief summary of the essential details of this formalism referred to here.

It is important to note that effective systems based on quantum moments are
more general than what is usually considered as a local effective
action. Derivative expansions of effective equations do not necessarily exist
for all quantum states, which for moment equations corresponds to an adiabatic
approximation (see below). Especially the algebra of effective constraints is
insensitive to the existence of such regimes or expansions. All one needs is
an expansion by moments which in principle could go to rather high orders. If
there is a consistent quantum system with a first-class algebra of constraint
operators $\hat{C}_I$, it follows from the methods introduced in
\cite{EffCons,EffConsRel} that there is a consistent algebra of effective
constraints $\langle\hat{C}_I\rangle$ as well as higher-order ones of the form
$\langle\widehat{\rm pol}\hat{C}_I\rangle$: According to (\ref{Poisson}), we
have $\{\langle\hat{C}_I\rangle, \langle\hat{C}_J\rangle\}=
\langle[\hat{C}_I,\hat{C}_J]\rangle/i\hbar$ for all $I$ and $J$. If the
constraint operators are first class, $\{\langle\hat{C}_I\rangle,
\langle\hat{C}_J\rangle\}$ is therefore an expectation value of a combination
of constraints, which vanishes when the effective constraints are all zero. If
the effective constraints $\langle\widehat{\rm pol}\hat{C}_I\rangle$ are
expanded to some order in the moments, they are first class to within this
order.  Requiring closed effective constraints therefore does not pose
conditions in addition to what one must assume for any consistent quantum
theory. A closed algebra of effective constraints must exist in all regimes,
not just semi-classical or low-curvature ones; one must, however, expand to
higher orders in the moments as states become more quantum.

Under certain conditions, combining a semi-classical with an adiabatic
expansion, equations of motion for moments, ${\rm d}\Delta(q^ap^b)/{\rm
  d}t=\{\Delta(q^ap^b),\langle\hat{H}\rangle\}$, can be partially solved, upon
which coupling terms in expectation-value equations take on the form of higher
time derivatives \cite{EffAc,Karpacz,HigherTime}. Higher time derivatives
therefore result from canonical quantum gravity, and they must obey covariance
conditions because the moment terms they originate from must leave the
first-class nature of the effective constraints $\langle\hat{C}\rangle$
intact. These corrections are important for any theory of quantum gravity, and
they are not related to or replaced by holonomy modifications. The moments
provide additional degrees of freedom, with Poisson brackets derived from
(\ref{Poisson}) and their dynamics following from (\ref{dOdt}).

Considering these detailed constructions, it is clear that effective
calculations do not assume what phase-space structure they refer to. Poisson
brackets of all variables follow from quantum commutators via
(\ref{Poisson}). In most cases of effective constraint algebras found so far,
moment-terms are ignored to avoid their more-complicated algebra. One is then
dealing with holonomy modifications without considering higher time
derivatives. In section \ref{s:Iso}, we will comment on how meaningful
this separation of holonomy modifications and higher-curvature (or moment)
corrections is.  What is left of the phase space is then just the expectation
values of basic operators, with Poisson brackets identical to the classical
ones. This fact may give rise to the impression that the classical phase-space
structure is just assumed, but it is actually implied by general effective
methods together with the order of calculations.

Sometimes, following \cite{Bohr}, one uses a partial derivation of effective
equations \cite{Taveras,EffRecollapse} as a shortcut, in which one assumes the
evolving state and its moments to be of a certain form, most often Gaussian
for simplicity. In \cite{Bohr}, the method has been introduced and used in
order to show that the correct classical limit is obtained for loop quantum
cosmology, building on \cite{SemiClass}. But it turned out to be less reliable
for a derivation of semi-classical physics with leading-order quantum
corrections: In most quantum systems, an initial Gaussian state rapidly
evolves away from being Gaussian (see for instance \cite{HigherMoments} in
quantum cosmology). Nevertheless, there may still be interesting features
shown by such models. In order to ensure that the moments $\Delta$ kept in the
system are in some sense close to an evolving state, one must find a relation
$\Delta(\langle\hat{q}\rangle,\langle\hat{p}\rangle)$ between them and
expectation values, so that a state peaked on
$(\langle\hat{q}\rangle(t),\langle\hat{p}\rangle(t))$ as it evolves according
to the Schr\"odinger equation changes its moments by $\Delta(t)=
\Delta(\langle\hat{q}\rangle(t),\langle\hat{p}\rangle(t))$. (For comparison,
in general effective equations such a relation is available only in special
regimes, such as adiabatic ones. Otherwise, there are coupled but independent
equations for
$(\langle\hat{q}\rangle(t),\langle\hat{p}\rangle(t),\Delta(t))$.) It is then
possible to write the system of
$(\langle\hat{q}\rangle,\langle\hat{p}\rangle,\Delta)$ as one on a phase space
of classical variables $(q,p)$, with Poisson brackets given by the pull-back of
the symplectic structure on state space by the embedding $(q,p)\mapsto
(\langle\hat{q}\rangle,\langle\hat{p}\rangle,\Delta)$. Since moments $\Delta$
contribute non-trivially to the full symplectic structure, the Poisson
brackets for $q$ and $p$ derived in this way may differ from the classical
one, and therefore carry quantum corrections. But such corrections follow from
one's choice of state, intermingling moments with basic expectation values;
they do not arise when one considers moments as independent degrees of
freedom, as one must do in general regimes. Moreover, in order to find the
embedding $(q,p)\mapsto (\langle\hat{q}\rangle,\langle\hat{p}\rangle,\Delta)$
one solves some equations of motion. In the present context of non-trivial
constrained systems, such an embedding, as well as the whole procedure it
gives rise to, is not available before a consistent algebra has been
derived. It is feasible only in homogeneous models which have just a single
constraint with a trivial Abelian algebra. For non-trivial effective
constraint algebras, on the other hand, equations of motion cannot be used to
derive the phase-space structure. Moments must be treated as independent of
expectation values, and the Poisson brackets between expectation values of
basic operators do not carry quantum corrections. This result is derived from
the quantum theory and not postulated.

\subsubsection{Inverse-triad corrections}

In addition to these two types of corrections, holonomy modifications and
quantum back-reaction which both depend on curvature parameters, we have
inverse-triad corrections from loop quantum gravity. These corrections have
been computed in several Abelian models
\cite{InvScale,Ambig,QuantCorrPert,InflTest} (using U(1) instead of SU(2)) and
evaluated numerically for SU(2) states \cite{BoundFull}. Their construction
shows that inverse-triad corrections depend on the relation of elementary flux
values (or the spatial discreteness scale) to the Planck scale. More
precisely, they refer to expectation values of elementary flux operators in a
quantum-gravity state. For a highly refined state, these values are
small. Rather surprisingly, inverse-triad corrections disappear for large flux
values, where discretization effects of spatial derivatives (and, presumably,
curvature terms) would become large. A parametrization of inverse-triad
corrections is then often expanded as
\begin{equation}
\langle\hat{E}\rangle\langle\widehat{E^{-1}}\rangle \sim 1+ c \frac{\ell_{\rm
  P}^2}{|F|}+\cdots\,,
\end{equation}
where $F$ is some flux or area value characteristic of the regime considered.
Inverse-triad corrections are therefore important in regimes different from
those where holonomy modifications or higher-curvature terms are
crucial. These corrections can easily be separated from each other, allowing
simplifications in the organization of calculations.

\subsubsection{Holonomy corrections and their ambiguities}

In an unexpanded form, holonomy modifications result from matrix elements of
SU(2)-holonomies. In most reduced models that have been studied in detail, a
reduction from SU(2) to U(1) comes along with the imposition of spatial
symmetries \cite{SymmRed,IsoCosmo} or diagonalization in Bianchi models
\cite{HomCosmo}. Holonomy modifications then take a form, such as
\begin{equation}\label{ume}
h(A)\sim \delta^{-1}\sin(\delta A)\,,
\end{equation}
of almost-periodic functions of the connection components
\cite{Bohr,IsoCosmo}. The parameter $\delta$, representing the lattice
structure, remains undetermined as long as the reduced model is not derived
from a consistent full theory. It is often assumed to be of Planck size, but
this assumption is not well-supported. First, $\delta$ comes from the
coordinate length of a curve, which by itself does not provide any physical
length scale. Secondly, if one takes elementary fluxes as an estimate of
$\delta$, inverse-triad corrections are large for Planckian values, implying
strong deviations from classical geometry \cite{Consistent}. At the present
state of knowledge, it is best to keep $\delta$ as an undetermined parameter,
which may perhaps be restricted by consistency conditions for a first-class
algebra of constraints or by future better information about a derivation from
the full theory.

In addition to the unknown magnitude of $\delta$, its possible dependence on
the triad is not much restricted. A $\delta$ depending on the triad amounts to
lattice refinement \cite{InhomLattice,CosConst}: the size of curves in a
discrete state changes as a space-time region expands or contracts. In the
full theory, the analog of a changing $\delta$ is the creation of new vertices
or curves as a discrete state evolves, which happens in current constructions
of the full dynamics. Again, it is the lack of a direct derivation from the
full theory that leaves ambiguities in suitable parametrizations of holonomy
modifications.

Currently, the closest to a derivation of symmetric models from the full
theory is a distributional construction of symmetric states and basic
operators \cite{SymmRed}. As recent considerations have shown \cite{NonAb},
the full quantum state space cannot be captured completely by a simple reduced
U(1)-theory. It is therefore unclear if U(1)-matrix elements such as (\ref{ume}), even with unspecified and triad-dependent
$\delta$, are sufficient to model the full expressions of the quantum
constraint. The almost-periodic form may be good as a first approximation, but
more generally one should try to keep the functional form of $h(A)$ as
unrestricted as possible. As shown especially by spherically symmetric models,
summarized in section \ref{s:Examples}, interesting results can still be found.

\subsubsection{Relation to homogeneous quantum cosmology (mini-superspace models)}

Before we continue with a more detailed consideration of the constraint
algebra, it is useful to recall the steps undertaken to formulate homogeneous
models of loop quantum cosmology. As in the full theory, one starts by
modifying the classical constraint so as to make it representable by
holonomies. The resulting expression is then quantized straightforwardly, just
inserting holonomy and flux (or inverse-flux) operators. When choosing the
modification (or the quantization if the modification is done only
  implicitly), one is subject to the same ambiguities, choices and unknowns
as just described for inhomogeneous models, and the resulting state equations
for wave functions are no less ambiguous than the effective constraints
including holonomy modifications.

After formulating the state equation, one has different options for a further
analysis. One may just solve the partial differential or difference equation
for the wave function, but then still has to compute expectation values of
suitable observables to derive physical implications. Effective methods
applied to homogeneous models provide a shortcut by which one can analyze the
same state equation but without computing a complete wave function. Being
based on expectation values and moments in their canonical incarnation,
effective methods directly produce results for observable quantities.

In the inhomogeneous context, effective methods for constrained systems
realize a further simplification, as already mentioned. For state equations,
one would have to show that constraint operators obey a first-class algebra
before they can consistently be solved. With the effective approach used here,
consistency is checked in terms of Poisson brackets, and once it is realized,
equations of motion for observables can directly be computed. In summary,
effective and standard constructions of models of quantum cosmology and their
solutions for wave functions are based on the same assumptions as far as their
relation to some full quantum gravity is concerned, and they are subject to
the same ambiguities. The difference lies in the approximations they employ to
derive solutions.

\section{Constraint algebra and space-time structure}

With the three types of corrections discussed in the preceding section (moment terms, holonomy modifications, and inverse-triad corrections) we
arrive at a modified constraint $C$. For consistency, it must be part of a
first-class algebra with the standard $\vec{D}$, so that equations of motion
and observables can be derived without conflicting statements. 

\subsection{Isolated modifications}
\label{s:Iso}

In general, all three types of corrections should be included in a modified
constraint of effective loop quantum gravity. However, we have already seen
that inverse-triad corrections depend on different quantities than holonomy
modifications and moment terms, and that they are in general relevant in
different regimes. It is therefore meaningful to study these corrections in
separation from the others, simplifying the system for a further analysis as
done in \cite{ConstraintAlgebra}. These corrections may then have potential
observational consequences even in regimes that are far from Planckian
curvature \cite{LoopMuk,InflTest,InflConsist}. 

When curvature becomes large, however, moment terms and holonomy modifications
must be included. It is not easy to separate these types of corrections
referring only to their dependence on the density or other parameters that
characterize the evolution of a state. Nevertheless, it is possible to
separate the computation of consistent versions with holonomy modifications on
the one hand, and moment terms on the other, and still infer reliable features
of space-time structure. Holonomy modifications change the dependence of the
effective constraint on expectation values of the connection, compared with
the classical expression. Independently of the specific form of this
modification, the presence of higher-order terms and higher spatial
derivatives of the connection is a general feature. Moment terms, on the other
hand, introduce new quantum degrees of freedom and their couplings to
expectation values. They may be derived for the standard constraint, or one
already modified by holonomy expressions. In either case, the general feature
is the presence of moments in addition to expectation values. 

In order to derive a consistent system, we are then looking at Poisson
brackets of the original classical constraints with different types of
characteristic terms added to them. For canonical variables such as the
connection (or extrinsic curvature) and the densitized triad, the Poisson
bracket of a moment and an expectation value, derived using (\ref{Poisson}),
always vanishes. Moment terms contribute to the Poisson bracket of two
constraints only when at least one moment is included on both sides of the
bracket. Another general consequence of (\ref{Poisson}) is that the Poisson
bracket of two moments always produces terms containing at least one
moment. Moments therefore modify the Poisson bracket of two constraints by
adding moment terms to the classical (or modified) bracket. Any anomaly
possibly produced by holonomy modifications will therefore remain in the
system when moment terms are added; it cannot be canceled generically, owing
to the different forms of new terms. (By a detailed analysis of the structure
of effective constrained systems $\langle\widehat{\rm pol}\hat{C}_I\rangle$
introduced in \cite{EffCons,EffConsRel}, one can show that moment corrections
do not deform the constraint algebra of the $\langle\hat{C}_I\rangle$
  \cite{EffConsQBR}. This observation is consistent with the fact that
higher-curvature effective actions do not modify the classical algebra
(\ref{DD})--(\ref{HH}) \cite{HigherCurvHam}.)  A system of constraints
containing holonomy modifications and moment terms can be consistent only when
the holonomy-modified version alone already produces a consistent version. And
any potential quantum correction in the constraint algebra of
holonomy-modified constraints implies a correction in the algebra that
contains also moment terms. It is therefore meaningful to study holonomy
modifications in isolation, even though numerically they are expected to be of
similar size as moment terms or higher-curvature corrections. Several
cosmological analyses have already been performed in this vein
\cite{ScalarHol,ScalarHolMuk,ScalarTensorHol}.

In the interplay of holonomy modifications and moment terms just described,
one could have cancellations only if the values of moments are tightly related
to expectation values, so that moment-independent anomalies could be canceled
by moment-dependent terms. In this case, the state would be restricted by
sharp relations between moments and expectation values. Formally, a consistent
version of constraints would result, and the restrictions on states may even
be of interest in early-universe cosmology, considering possible derived
initial conditions. (But one would have to make sure that the restricted
states are still consistent with the correct semi-classical limit of the
theory.) However, such effective constraints would not belong to a consistent
quantum theory because the requirement of a first-class algebra (and not just
consistency of the constraint equations themselves) would restrict the states.
This situation would be similar to the one encountered when anomalous
constraint operators overconstrain states by 
\[
[\hat{O}_1,\hat{O}_2]\psi= \hat{O}_1\hat{O}_2\psi- \hat{O}_2\hat{O}_1\psi=0\,,
\]
which is a condition independent of $\hat{O}_1\psi=0=\hat{O}_2\psi$
for constraint operators that are not first class.

\subsection{Examples of anomaly-free constraints}
\label{s:Examples}

Consistent deformations of the classical algebra have been found in two
classes of models, spherically symmetric ones and cosmological
perturbations. In order to compare these results, we include here brief
summaries; see \cite{JR,HigherSpatial,ScalarHolInv} for details.

\subsubsection{Spherical symmetry}

For spherical symmetry, a general Hamiltonian constraint can be written as
\begin{eqnarray} \label{HQsph}
C^{\rm Q}_{\rm grav}[N]&=&-\frac{1}{2G}\int {\rm d} x\,
N\bigg[\alpha|E^x|^{-\frac{1}{2}} E^{\varphi}f_1(K_{\varphi},K_x)+
2\bar{\alpha} |E^x|^{\frac{1}{2}}f_2(K_{\varphi},K_x) \nonumber\\ 
&&\qquad +\alpha_{\Gamma}|E^x|^{-\frac{1}{2}}(1-\Gamma_{\varphi}^2)E^{\varphi}+
2\bar{\alpha}_{\Gamma}\Gamma_{\varphi}'|E^x|^\frac{1}{2} \bigg]\,,
\end{eqnarray}
a function of two pairs of canonical fields $(K_x,E^x)$ and
$(K_{\varphi},E^{\varphi})$ where
$\Gamma_{\varphi}=-(E^x)'/2E^{\varphi}$. (See \cite{SphSymm} for these
spherically symmetric variables.)  This form includes inverse-triad
corrections if the $\alpha$'s are not all equal to one, as well as (pointwise)
holonomy corrections via two new functions $f_1$ and $f_2$. Holonomy
corrections should also give rise to higher spatial derivatives if the curve
integration is expanded, but the status of such terms regarding consistent
constraint algebras remains incomplete \cite{HigherSpatial}. Classically,
$f_1=K_{\varphi}^2$ and $f_2=K_xK_{\varphi}$, as a reduction of (\ref{CA})
shows.

Anomaly freedom can be realized if $f_2=K_x F_2(K_{\varphi},E^x,E^{\varphi})$
provided that \cite{JR}
\begin{equation} \label{alphaGamma}
 \left[\bar{\alpha}\alpha_{\Gamma} -2E^x
   \left(\frac{\partial\bar{\alpha}}{\partial      E^x}\bar{\alpha}_{\Gamma}-
 \bar{\alpha}\frac{\partial\bar{\alpha}_{\Gamma}}{\partial E^x}\right)\right]
F_2+ 2\bar{\alpha}\bar{\alpha}_{\Gamma} E^x \frac{\partial F_2}{\partial E^x}=
\frac{1}{2}\alpha\bar{\alpha}_{\Gamma} \frac{\partial f_1}{\partial K_{\varphi}}\,.
\end{equation}
(Since $f_2$ must depend linearly on $K_x$, no holonomy modifications for
$K_x$ can be implemented in this way, unless perhaps one includes higher
spatial derivatives.) The deformation function in (\ref{HHbeta}) is then
\begin{equation} \label{betasph}
 \beta_{\rm sph}=\bar{\alpha}\bar{\alpha}_{\Gamma} \frac{\partial
F_2}{\partial K_{\varphi}}\,.
\end{equation}

The holonomy-modification function $F_2$ primarily depends on $K_{\varphi}$,
but it may depend on the triad components as well through lattice
refinement. However, if $F_2$ is almost periodic in $\delta(E^x,E^{\varphi})
K_{\varphi}$ as expected for Abelian holonomies, $\partial F_2/\partial E^x$
is not almost periodic in $K_{\varphi}$, and (\ref{alphaGamma}) is not
compatible with an almost periodic $f_1$. Lattice refinement can therefore not
be implemented in these models, and $F_2$ cannot depend on $E^x$ if one
insists on almost-periodic holonomy modifications. (It is interesting to
compare this conclusion with the results of \cite{SIGMA}, a construction of
homogeneous non-Abelian models which naturally lead to lattice refinement but
are not based on almost-periodic functions.) With this condition, we can solve
(\ref{alphaGamma}) for
\begin{equation} \label{F2long}
 F_2= \frac{1}{2}\frac{\partial f_1}{\partial K_{\varphi}}
 \frac{\alpha\bar{\alpha}_{\Gamma}}{\bar{\alpha}\alpha_{\Gamma}-2E^x
   \left(\bar{\alpha}_{\Gamma}\partial\bar{\alpha}/\partial      E^x-
 \bar{\alpha}\partial\bar{\alpha}_{\Gamma}/\partial E^x\right)} \,.
\end{equation}
Since inverse-triad correction functions do depend on $E^x$, the
$E^x$-independence of $F_2$ implies that
\begin{equation} \label{alphaGamma2}
\alpha\bar{\alpha}_{\Gamma}-\bar{\alpha}\alpha_{\Gamma}+2E^x
   \left(\bar{\alpha}_{\Gamma}\frac{\partial\bar{\alpha}}{\partial E^x}-
 \bar{\alpha}\frac{\partial\bar{\alpha}_{\Gamma}}{\partial E^x}\right)=0
\end{equation}
and $F_2$ simplifies to
\begin{equation} \label{F2}
 F_2= \frac{1}{2}\frac{\partial f_1}{\partial K_{\varphi}}\,.
\end{equation}
(The same relation can be found using operator methods \cite{SphSymmOp}.)
The deformation function is then
\begin{equation} \label{betasph2}
\beta_{\rm sph}= \frac{1}{2}\bar{\alpha}\bar{\alpha}_{\Gamma}
\frac{\partial^2 f_1}{\partial K_{\varphi}^2}\,.
\end{equation}

\subsubsection{Cosmological perturbations}
\label{sss:Cosmpert}

For perturbative treatments, one writes the variables as
$K_a^i=\bar{k}\delta_a^i+\delta K_a^i$ and $E^a_i=\bar{p}\delta^a_i+ \delta
E^a_i$. (See section \ref{s:Pert} below for further details on the perturbation
scheme.) The expanded Hamiltonian constraint can then be parametrized as
\begin{equation}
C^{\rm Q}_{\rm grav}[N] = \frac{1}{8\pi G} \int_{\Sigma} {\rm d}^3x \left[
  \bar{N}(\mathcal{H}^{(0)}+\mathcal{H}^{(2)})+\delta N
  \mathcal{H}^{(1)}\right]\,, 
\label{HolModHam}
\end{equation}
where 
\begin{equation}\label{HQcosmo0}
\mathcal{H}^{(0)} = -6 \gamma_0 \sqrt{\bar{p}} (\mathbb{K}[1])^2
\end{equation}
contains pointwise holonomy corrections assumed to be of the specific form
${\mathbb K}[1]=\delta^{-1}\sin(\delta \bar{k})$ and inverse-triad corrections
via a function $\gamma_0$. The first- and second-order contributions
\begin{equation} \label{HQcosmo1}
\mathcal{H}^{(1)}  = -4\gamma_0\sqrt{\bar{p}} \left(  \mathbb{K}[s_1]
  +\alpha_1\right) 
\delta^c_j \delta K^j_c-\frac{\gamma_0}{\sqrt{\bar{p}}} 
\left(  \mathbb{K}[1]^2+\alpha_2  \right) \delta^j_c\delta E^c_j 
+\frac{2\gamma_0}{\sqrt{\bar{p}}} (1+\alpha_3) \partial_c \partial^j \delta E^c_j
\end{equation}
with $\mathbb{K}[s]= (s\delta)^{-1}\sin(s\delta\bar{k})$, and
\begin{eqnarray}
\mathcal{H}^{(2)}  &=& \gamma_0\sqrt{\bar{p}}(1+\alpha_4) \delta K^j_c \delta
K^k_d \delta^c_k \delta^d_j   
-\gamma_0\sqrt{\bar{p}} (1+\alpha_5)(\delta K^j_c
\delta^c_j)^2 \label{HQcosmo2} \\ 
&&-\frac{2\gamma_0}{\sqrt{\bar{p}}} \left(  \mathbb{K}[s_2]+\alpha_6 \right)
\delta E^c_j \delta K^j_c    
-\frac{\gamma_0}{2\bar{p}^{3/2}} \left( \mathbb{K}[1]^2+\alpha_7\right)\delta
E^c_j \delta E^d_k \delta^k_c \delta^j_d \nonumber \\ 
&&+\frac{\gamma_0}{4\bar{p}^{3/2}}\left( \mathbb{K}[1]^2+\alpha_8\right) (\delta
E^c_j \delta^j_c)^2  
- \frac{\gamma_0}{2\bar{p}^{3/2}}(1+\alpha_9) \delta^{jk} (\partial_c \delta
E^c_j)(\partial_d \delta E^d_k) \nonumber  
\end{eqnarray}
contain the same functions but also ``counterterms'' $\alpha_i$ inserted for
sufficient freedom to incorporate consistent modifications. Also here, terms
with higher spatial derivatives are expected but have not been implemented
yet.  Anomaly-free versions \cite{ScalarHol,ScalarHolInv} impose several
conditions on the counterterms but leave $\gamma_0$, $\alpha_3$ and ${\mathbb
  K}[1]$ free. (The form ${\mathbb K}[1]=\delta^{-1}\sin(\delta \bar{k})$
quoted above has been chosen to agree with suggestions from mini-superspace
models, but it does not follow from the condition of anomaly-freedom.)  They
lead to a deformation function
\begin{equation}\label{betacosmo}
\beta_{\rm cosmo}=\frac{1}{2}\gamma_0 (1+\alpha_3)
\frac{\partial^2(\gamma_0{\mathbb K}[1]^2)}{\partial\bar{k}^2}
\end{equation}
for pure gravity. A matter Hamiltonian would have its own correction
functions that can contribute factors to $\beta_{\rm cosmo}$.

An explicit example of anomaly cancellation in the presence of
inverse-volume corrections is reported in Appendix \ref{App:Consist}.

\subsubsection{Comparison} \label{323}

At first sight, the treatments in those two classes of models look rather
different, owing to the presence of counterterms only in the perturbative
approach. Existing treatments of spherically symmetric and cosmological
  models have indeed appeared to be rather separate from each other. However,
the final results exhibit nice qualitative agreements. Our detailed
  comparison of the relevant terms allows us to develop a unified picture of
  anomaly-free constraints, which we do in the remainder of this
  subsection. One of the consequences is that the ``counterterms'' used in
  cosmological models, sometimes with a flavor of arbitrariness, are nothing
  but different parametrizations of quantization ambiguities as they also
  appear in spherically-symmetric models.

As far as the inverse-triad contributions are concerned, the two factors
$\bar{\alpha}\bar{\alpha}_{\Gamma}$ in (\ref{betasph2}) come from one term in
(\ref{HQsph}) quadratic in extrinsic curvature and one term with second-order
spatial derivatives of the densitized triad. The two factors of $\gamma_0$ and
$\gamma_0(1+\alpha_3)$ in (\ref{betacosmo}) come from similar terms in the
perturbed Hamiltonian constraint, the term quadratic in extrinsic curvature in
(\ref{HQcosmo0}) and a term with second-order spatial derivatives of the
densitized triad in (\ref{HQcosmo1}), respectively. Even though the models as
well as the calculations that lead up to a consistent constraint algebra are
rather different, the forms of their deformation functions $\beta_{\rm sph}$ in
(\ref{betasph}) and $\beta_{\rm cosmo}$ in (\ref{betacosmo}) are the same. For
instance, for the common choice of almost-periodic holonomy modifications,
such that $f_1(K_{\varphi})=\delta^{-2}\sin^2(\delta K_{\varphi})$ and
${\mathbb K}[1]^2=\delta^{-2}\sin^2(\delta \bar{k})$, we have $\beta_{\rm
  sph}=\cos(2\delta K_{\varphi})$ and $\beta_{\rm cosmo}=\cos(2\delta\bar{k})$
in the absence of inverse-triad corrections. These correction functions are
negative at around the maximum of holonomy modifications, an interesting
feature that implies signature change; see section \ref{s:Sig}.

As is clear from these comparisons, the counterterms in the perturbative
treatment play a role just like the different correction functions in the
parametrization for spherically symmetric models. This observation suggests a
revised interpretation of counterterms. In \cite{ConstraintAlgebra},
counterterms were introduced as modifications of the classical constraint that
appear to be necessary for an anomaly-free algebra but, unlike primary
correction functions $\gamma_0$, could not be guessed easily from known
features of inverse-triad or holonomy operators. Moreover, the classical
compact form of the full Hamiltonian constraint, in which inverse triads
appear in a single factor of $\epsilon^{ijk}E^a_iE^b_j/\sqrt{|\det E|}$,
suggested that there should be just one total correction factor for the
Hamiltonian constraint in the presence of inverse-triad corrections. Since
such a single factor did not lead to an anomaly-free constraint, the
$\alpha_i$ were therefore introduced to ``counter'' anomalies that a uniform
factor of $\gamma_0$ would imply.

As shown in spherically symmetric models with their four types of $\alpha$, it
is more straightforward, and perhaps less confusing, to view all modified
terms on the same footing. They would all come from inverse-triad corrections
or from a combination with holonomy corrections, but different terms in the
Hamiltonian constraint would require different modifications. The presence of
several different terms merely shows that consistent modifications are not as
simple as initially expected, and the freedom in inserting them in the
constraint reflects general quantization ambiguities. Lacking a derivation of
effective constraints from a full operator (as well as consistent versions of
full operators), the form of modification functions cannot be determined
without further input. Fortunately, however, the requirement of anomaly
freedom implies conditions strong enough to fix some (but not all) of the free
functions. (There are two types of conditions for anomaly freedom: those
coming from the closure of $\{C[N],\vec D[\vec M]\}$ and those from
$\{C[N_1],C[N_2]\}$. The first condition constitutes simple geometrical
consistency of spatial tensors: for instance, the correction functions in
$C[N]$ should yield the appropriate density weight, and all indices should be
properly contracted.  The latter condition, referring to space-time geometry,
is much more non-trivial to satisfy and can help to rule out some quantization
ambiguities.)

\subsubsection{Free functions in parametrizations}\label{fref}

After imposing anomaly freedom, several functions remain free. In spherically
symmetric models, one of the two holonomy modification functions, $F_2$, is
fully determined in terms of the other correction functions.  Moreover, the
form of these relations indicates further restrictions. For instance, because
holonomy modification functions $F_1$ and $F_2$ should have a characteristic
form rather different from the inverse-triad correction functions $\alpha$ and
$\alpha_{\Gamma}$, as discussed, (\ref{F2long}) implies
(\ref{alphaGamma2}). Only three of the inverse-triad correction functions then
remain free. (One can think of this freedom as independent quantization
ambiguities in the quantization of the explicit $|E^x|^{-1/2}$ in
(\ref{HQsph}) as well as different functions of inverse triad components
$1/E^{\varphi}$ in the spin connection $\Gamma_{\varphi}$.)

\subsection{Perturbation scheme}
\label{s:Pert}

In cosmological models, one does not perform calculations of consistent
constraint algebras for the fields $E^a_i$ and $A_a^i$ (or $K_a^i$), but for
variables split into background quantities and inhomogeneous
perturbations. This procedure is no different from some other calculations in
cosmology, especially regarding the early universe. But, again following
\cite{ConstraintAlgebra,ScalarGaugeInv} (see also \cite{PertObsI,PertObsII}),
there are several special constructions required for the canonical methods
used to derive constraint algebras.

Assuming some inhomogeneous field $q(x)$, we split it into a background part
$\bar{q}$ and its inhomogeneity $\delta q(x)$ by writing $q(x)=\bar{q}+\delta
q(x)$.\footnote{So far, the background for perturbative calculations in
  models of loop quantum cosmology has been isotropic and spatially flat. Our
  considerations apply also to general homogeneous backgrounds, as they would
  be required for anisotropic back-reaction.} For most calculations, $\delta
q$ will be assumed small in cosmology, but for now the splitting is just a
different parametrization of field degrees of freedom. We only require
$\int{\rm d}^3x\, \delta q(x)=0$ in order to avoid overcounting of the
homogeneous variables. In other words, $\delta q$ is a pure inhomogeneity.

If we start with general canonical variables $(q,p)$, the decompositions
$q=\bar{q}+\delta q$ and $p=\bar{p}+\delta p$ map them into a larger phase
space $\bar{\cal P}$ with variables $(\bar{q},\bar{p};\delta q,\delta p)$. The
map is not surjective unless one requires $\int{\rm d}^3x\, \delta q(x)=0$ and
$\int{\rm d}^3x\, \delta p(x)=0$ on $\bar{\cal P}$, which can be done by
constraints. Before these linear constraints are imposed, we have the standard
Poisson brackets $\{\bar{q},\bar{p}\}_{\bar{\cal P}}=1/V_0$ and $\{\delta
q(x),\delta p(y)\}_{\bar{\cal P}}=\delta(x,y)$ for two independent sets of
variables, $(\bar{q},\bar{p})$ for the background and fields $(\delta q,\delta
p)$.

In order to avoid overcounting degrees of freedom, the phase space $\bar{\cal
  P}$ must be constrained by imposing $C_q=\int{\rm d}^3x\, \delta q(x)=0$ and
$C_p=\int{\rm d}^3x\, \delta p(x)=0$. If these integrations are performed over
a finite but sufficiently large region of coordinate volume $V_0=\int{\rm
  d}^3x$, we have $\{C_q,C_p\}_{\bar{\cal P}}=V_0$.  The constraints are
therefore second-class constraints, and they ensure that $\bar{q}$ and
$\bar{p}$ are indeed spatial averages: $\bar{q}=V_0^{-1}\int{\rm d}^3x\, q(x)$
and $\bar{p}=V_0^{-1}\int{\rm d}^3x\, p(x)$.

The canonical structure of $q$ and $p$ follows from the Liouville term
$\int{\rm d}^3x\, \dot{q}p$ in the canonical action. By inserting our
decompositions (or pulling back the symplectic form), we find 
\begin{equation}
 \int{\rm d}^3x\, \dot{q}p= \int{\rm d}^3x\, (\dot{\bar{q}}+\delta
 \dot{q})(\bar{p}+\delta p)= V_0 \dot{\bar{q}}\bar{p}+ \int{\rm d}^3x\,
 \delta\dot{q}\delta p\,,
\end{equation}
using the constraints $C_q=0$ and $C_p=0$. If we work on the unrestricted
phase space $\bar{\cal P}$ on which $C_q=0$ and $C_p=0$ are not solved
explicitly, the presence of second-class constraints is taken into account by
using the Dirac bracket
\begin{equation}
 \{\delta q(x),\delta p(y)\}= \delta(x,y)-\frac{1}{V_0}\,.
\end{equation}
With the subtraction of $1/V_0$, integrating over space on both sides, in
either $x$ or $y$, produces zero. For the background variables, we have the
homogeneous Poisson bracket
\begin{equation}
\{\bar{q},\bar{p}\}= \frac{1}{V_0}\,.
\end{equation}

We now assume the inhomogeneity to be small and derive equations of motion.
For equations of first order in inhomogeneity, we expand the constraints
generating evolution to second order:
\begin{equation}
 C[N]=\int{\rm d}^3x\, NC = \int{\rm d}^3x\, (\bar{N}\bar{C}+ \delta N C^{(1)}+
 \bar{N} C^{(2)}) 
\end{equation}
(plus contributions from the diffeomorphism constraint). The superscript
indicates the order in inhomogeneity, and we have used a decomposition
$N=\bar{N}+\delta N$ as in the canonical variables. Taking Poisson brackets
with $\delta q$, for instance, produces a first-order term from $C^{(2)}$,
which thus contributes to first-order equations of motion.

In more detail, we have four types of equations: constraints
\begin{equation}
 \bar{C}+C^{(2)}=0\quad\mbox{and}\quad C^{(1)}=0
\end{equation}
from varying by $\bar{N}$ and $\delta N$,
and equations of motion
\begin{eqnarray}
\dot{\bar{q}}=\{\bar{q},\bar{C}[\bar{N}]+C^{(1)}[\delta N]+
C^{(2)}[\bar{N}]\}\,,\quad &&\quad
\dot{\bar{p}}=\{\bar{p},\bar{C}[\bar{N}]+C^{(1)}[\delta N]+ 
C^{(2)}[\bar{N}]\}\,,\label{EvolveBackground}\\
\delta\dot{q}= \{\delta q,C^{(1)}[\delta
N]+C^{(2)}[\bar{N}]\}\,,\quad&&\quad\delta\dot{p}= \{\delta p,C^{(1)}[\delta
N]+C^{(2)}[\bar{N}]\}\,, \label{EvolveMode} 
\end{eqnarray}
with $\bar{C}[\bar{N}]= V_0\bar{N}\bar{C}$, $C^{(2)}[\bar{N}]=\int{\rm d}^3x\,
\bar{N}C^{(2)}$, and so on. In perturbative treatments, the background lapse
function $\bar{N}$ is usually fixed.  In this way, a clear interpretation of
evolution for modes on a given background results. Fixing $\bar{N}$ thereby
determines the background time.

The backreaction term $b:=\{\bar{q},C^{(1)}[\delta N]+ C^{(2)}[\bar{N}]\}$ in
the background equation of motion does contain $\delta N$, but it is
independent of the gauge choice.  To see this, we note that the modes
themselves can be combined to gauge-invariant variables $\phi$ by ensuring
that $\{\phi,C^{(1)}[\epsilon]\}=0$ for all $\delta N=\epsilon$ (while
$\dot{\phi}=\{\phi,\bar{C}[\bar{N}]+C^{(2)}[\bar{N}]\}\not=0$ for evolving
quantities if the background gauge is fixed). For the back-reaction term, we
then have $\{b,C^{(1)}[\epsilon]\}=0$ if the constraints are first class so
that $\{C[\bar{N}+\delta N],C[0+\epsilon]\}=0$ when the field equations are
satisfied. (We use the Jacobi identity in this statement, as well as the fact
that the background term $\{\bar{q},\bar{C}[\bar{N}]\}$ Poisson commutes with
$C^{(1)}[\epsilon]$.) The back-reaction term is therefore a combination of
gauge-invariant perturbations, provided the constraint algebra is first class.

In these equations, the second-order constraint is required to generate
first-order equations for the modes, but it also contributes back-reaction to
the background variables in (\ref{EvolveBackground}). Usually, this term is
small for early-universe cosmology and can be ignored when equations of motion
are solved. (It cannot be ignored at the stage of constructing a consistent
gauge system, for without $C^{(2)}$ in (\ref{EvolveBackground}), but still in
(\ref{EvolveMode}), the set of equations is not Hamiltonian, making it much
more difficult to check whether the system is anomaly-free.) However, there
are situations, especially derivations of non-Gaussianity, where one must go
to higher orders. The perturbed equations of motion are then non-linear even
for small $\delta q$, and may be difficult to solve. An alternative
perturbation scheme is used in this context, in which the variables are
expanded (not decomposed) as $q=q_{\rm B}+\delta^{[1]}q+\delta^{[2]}q+\cdots$,
where $q_{\rm B}$ is a background zero-order quantity.  To each order, linear
equations of motion result if lower-order solutions have already been
obtained. Such an expansion is therefore more convenient for solving the
equations, but it does not give rise to a well-defined symplectic structure
for the independent orders (the $\delta^{[n]}q$ are not independent degrees of
freedom for different $n$).  Therefore, it can be used only at the level of
equations of motion, but not when consistent algebras and Poisson brackets are
still to be computed. For the latter, the well-defined symplectic structure of
$(\bar{q},\bar{p};\delta q,\delta p)$ is essential.

The relation between these two types of expansions is not straightforward. Up
to linear order, we have $\bar{q}=q_{\rm B}$ and $\delta q=\delta^{[1]}q$. But
with higher orders, even the background variables do not necessarily agree
because $\int{\rm d}^3x\, (\delta^{[1]}q+\delta^{[2]}q+\cdots)\not=0$ in
general. Moreover, $\bar{q}$ and $q_{\rm B}$ are subject to different
equations, $q_{\rm B}$ to the background equation $\dot{q}_{\rm
  B}=\{\bar{q},\bar{C}[\bar{N}]\}|_{\bar{q}=q_{\rm B}}$ independent of any
inhomogeneity, while the equation for $\dot{\bar{q}}$ contains a back-reaction
term $\{\bar{q},C^{(2)}[\bar{N}]\}$. In general, $\bar{q}$ and $q_{\rm B}$ are
therefore different, and $\delta q\not=\delta^{[1]}q+\delta^{[2]}q+\cdots$.
(The two possible schemes have been contrasted also in \cite{AAN}, but without
paying due attention to symplectic properties and the distinction between
independent degrees of freedom and higher perturbative orders.)  For higher
orders in inhomogeneity in situations in which a consistent constraint algebra
must first be found, one uses higher-order constraints $C=\int{\rm d}^3x\,
[\bar{N}(\bar{C}+C^{(2)}+C^{(3)})+\delta N(C^{(1)}+B^{(2)})]$ (without
$(\delta N)^2$ because the constraint is linear in $N$). As indicated,
expanding the constraint $C$ now gives rise to two different second-order
terms which we distinguish from each other by calling them $C^{(2)}$ and
$B^{(2)}$. Here, one inserts $q=\bar{q}+\delta q$ and $p=\bar{p}+\delta p$ in
the non-linear $C$ and expands in $\delta q$ and $\delta p$ to a desired
order. Superscripts indicate the order in $\delta q$ taken from
$q=\bar{q}+\delta q$; no $\delta^{[i]}$ is used at this stage since one still
needs the symplectic structure, which only exists for $(\bar{q},\bar{p};\delta
q,\delta p)$, to ensure consistency and derive equations of motion and gauge
transformations. If the algebra is consistent, one obtains the same type of
equations as before, but with additional terms: A background constraint
$\bar{C}+C^{(2)}+C^{(3)}$, which also generates background equations of motion
plus back-reaction when smeared with a background lapse $\bar{N}$, a
constraint $C^{(1)}+B^{(2)}$ for inhomogeneity, and their equations of motion
generated by $C^{(2)}[\bar{N}]+C^{(3)}[\bar{N}]+ B^{(2)}[\delta N]$.

If a consistent set of third-order constraints has been found, giving rise to
second-order equations of motion, one can proceed to solving them. (No such
explicit version is known yet. Second-order perturbation equations have been
considered in \cite{NonGaussInvVol}, but only using conditions found for
anomaly-free first-order equations.) At second order, the equations of motion
for $(\bar{q},\delta q)$ are non-linear, but they may now be reformulated for
further analysis. In particular, one can, at this stage only, transform to
second-order linearizing variables $(q_{\rm
  B},\delta^{[1]}q,\delta^{[2]}q)$ by inserting
\begin{equation} \label{qqB}
 \bar{q}=q_{\rm
  B}+\delta^{[1]}q+\delta^{[2]}q-\delta q
\end{equation} 
in the consistent equations for $\dot{\bar{q}}+\delta\dot{q}$. (Note that
$\bar{q}$ and $\delta q$ have the same evolution generator, even though some
terms may drop out as in (\ref{EvolveMode}) compared to
(\ref{EvolveBackground}).)  According to the definition of the decomposition
of $q(x)=\bar{q}+\delta q$ and the expansion $q=q_{\rm
  B}+\delta^{[1]}q+\delta^{[2]}q$, this equation, to the orders considered, is
an identity. It may not be obvious to see that it provides a set of equations
for $(q_{\rm B},\delta^{[1]}q,\delta^{[2]}q,\ldots)$ only, because initially
just $\bar{q}$ is eliminated while the $\delta q$ in (\ref{qqB}) must still
cancel those in the equation of motion. Such cancellations would be obvious if
the Hamiltonian would be known to arise from one for $q(x)$, so that inserting
different decompositions would just be a reformulation of variables. However,
consistent constraints for cosmological perturbation theory have so far been
found only at the perturbative level (and not even to third order yet) at
which quantum modifications are inserted and potential anomalies canceled only
after $q(x)$ has already been substituted by $\bar{q}+\delta q$. It would be
extremely complicated to arrange a consistent perturbative system to one that
extends to the original $q$. Nevertheless, one can see that the substitution
(\ref{qqB}) does lead to a set of equations for $(q_{\rm
  B},\delta^{[1]}q,\delta^{[2]}q,\ldots)$. One need only note that the $\delta
q$ cancel after such a substitution if the consistent perturbative system can
be written as a truncation of {\em some} constraint for an undecomposed
$q(x)$; for the cancellation to happen it is not necessary that this
$q(x)$-system be consistent or anomaly-free. Since all higher-order terms are
then completely free, such an extension easily exists and the transformation
of variables is possible. The new equations are linear if one proceeds order
by order in the new variables, and they are consistent to the orders
considered.

\subsection{Deformed general relativity}

While a deformation such as (\ref{HHbeta}) provides a consistent gauge system
with unambiguous physical observables decoupled from gauge degrees of freedom,
it cannot belong to a space-time theory in the classical sense. Any generally
covariant theory of fields on a Riemannian space-time has as (part of) its
gauge content generators that obey the classical hypersurface-deformation
algebra with (\ref{HH}), or (\ref{HHbeta}) with $\beta=\pm 1$ (the sign being
space-time signature). Gauge transformations corresponding to the algebra
(\ref{DD})--(\ref{HH}) are, on the constraint surface, in one-to-one
correspondence with space-time Lie derivatives along vector fields with
components determined by the smearing functions of $C$ and $\vec{D}$
\cite{DiracHamGR,LapseGauge}. Second-order field equations for the metric must
equal Einstein's \cite{Regained}, and therefore there is no room for quantum
modifications, such as holonomies, that do not imply higher time
derivatives. If the canonical gauge system with constraints obeying the
algebra (\ref{DD}) and (\ref{HD}) with (\ref{HHbeta}) is to be interpreted in
space-time terms, one cannot refer to classical Riemannian space-time. 

Such a consequence is not altogether surprising in a background-independent
quantum theory of gravity. In loop quantum gravity, one takes pains avoiding
any background metric structure in the definition of basic operators and the
dynamics, hoping that some space-time picture emerges in a semi-classical
limit. However, even if the correct semi-classical limit is realized, there is
no guarantee that in any realistic regime there will be no deviations from the
classical space-time structure. It is often assumed that the space-time
structure remains classical while the dynamics can change, ideally in an
observable way. However, the dynamics in a generally covariant theory is
closely related to the space-time structure, and therefore any assumption that
the latter remains unchanged is likely to be wrong. This statement is
substantiated by the current constructions of consistent
hypersurface-deformation algebras with modifications according to loop quantum
gravity, all of which lead to deformations of the space-time structure.

\subsubsection{Deformations}

All known consistent versions of effective constraints of loop quantum gravity
generically imply deformations of the algebra, with corrections not just of
the constraints but also of structure functions in the form (\ref{HHbeta}). In
the models studied so far, it is possible to have modifications of the
Hamiltonian constraint that respect the classical algebra, for instance for
spherically symmetric inverse-triad corrections (\ref{HQsph}) with
$\bar{\alpha}=\bar{\alpha}_{\Gamma}=1$ and $\alpha=\alpha_{\Gamma}\not=1$
\cite{LTBII}, or for suitable choices in cosmological models
\cite{ScalarHolInv}. However, there is no indication why such choices should
be preferred, and as they are very special, they do not give insights about
the generic situation. (Moreover, if the constraint algebra is classical, it
implies the classical dynamics for general, that is non-symmetric and
non-perturbative, geometries \cite{Regained}. The possibility of undeformed
constraint algebras with modified dynamics therefore seems to be a spurious
feature of the restricted situations studied so far. In fact, as shown in
\cite{Action}, the constraint algebra does not always uniquely determine the
dynamics unless one allows for general geometries.)

The deformations found appear to be universal: For inverse-triad
corrections, (\ref{HH}) is modified by the product of different types
of the primary inverse-triad correction functions of the form
$\alpha(\langle\hat{E}\rangle)=\langle\hat{E}\rangle
\langle\widehat{E^{-1}}\rangle$. These functions are always positive
and approach small values at small fluxes, where the ``smallness'' is
determined by the discreteness scale appearing in the
corrections. Such deformed algebras have been derived using effective
methods for cosmological perturbations \cite{ConstraintAlgebra} and in
spherically symmetric models \cite{JR,LTBII}, as well as operator
methods in $2+1$ Abelian models
\cite{TwoPlusOneDef,TwoPlusOneDef2,AnoFreeWeak} and spherically symmetric
models \cite{SphSymmOp}.

For holonomy corrections, the deformation (by a cosine function if the
holonomy modification is by the standard sine) appears at the same place, but now
depends on the connection or extrinsic curvature. In particular, this function is
not restricted to be positive, implying very strong modifications at high
curvature where the negative sign implies signature change \cite{Action}. More
generally, as shown by (\ref{betasph}) and (\ref{betacosmo}), the deformation
function $\beta$ is negative around the maximum of holonomy-modification
functions, being proportional to the second derivative. Such consistent
constraints have been derived in the same type of models, for cosmological
perturbations \cite{ScalarHol} and in spherically symmetric models \cite{JR}
using effective methods, and for operators in \cite{ThreeDeform}.

In both cases, the deformation function $\beta$ depends on spatial quantities
(the kinematical phase-space variables) and may appear to be slicing dependent
when interpreted in a conventional space-time picture. However, the fact that
the theory is defined by a closed constraint algebra implies that it presents
a fully consistent gauge system. There is no slicing dependence simply because
a deformed algebra (\ref{HHbeta}) belongs to a non-classical, non-Riemannian
space-time structure in which the phase-space functions are no longer the
standard spatial metric and extrinsic curvature. Thanks to anomaly freedom,
for every classical gauge transformation that would lead to a new slicing,
there is a quantum-corrected transformation in the deformed algebra. The
correct number of degrees of freedom is realized, and all auxiliary structures
(classically amounting to a space-time slicing) are removed. In physical
terms, the system is to be evaluated by computing observables, rather than
mathematical constructs such as slicings of space-time.  (In some cases,
interestingly, one can develop a standard space-time model by absorbing the
deformation function $\beta$ in the inverse metric by a canonical
transformation \cite{Absorb}. After the transformation, the constraint algebra
then is the classical one and generates coordinate transformations. But one
still does not obtain standard space-time geometry because the momenta of the
new metric components are not related to extrinsic curvature.)

As already discussed, the deformations found for holonomy modifications are
reliable even though moment terms have not yet been included. Moment terms
cannot cancel these deformations, but they could conceivably modify them.
However, no evidence for this exists at present. (Higher-curvature corrections
with their time derivatives do not deform the constraint algebra
\cite{HigherCurvHam}.)  Current calculations of the constraint algebra for
holonomy modifications are incomplete for another reason: Although they take
into account higher orders of the connection or extrinsic curvature, they do
not include higher spatial derivatives as suggested by a derivative expansion
of curve integrals in holonomies \cite{QuantCorrPert}. At high curvature,
higher spatial derivatives are generically as important as higher powers of
the connection; after all, Riemann curvature is a sum of quadratic
combinations of the connection and its first-order spatial
derivatives. Compared to inverse-triad corrections, the status of holonomy
modifications is therefore less complete. Nevertheless, the deformations they
imply appear to be reliable: Higher spatial derivatives would contribute terms
whose order of derivatives is invariant under taking Poisson brackets. (One
would just integrate by parts to rewrite expressions, preserving the number of
spatial derivatives.) Any anomaly produced by the higher-order terms of
holonomy modifications would therefore remain in the presence of higher
spatial derivatives. (Note the similarity of this statement to the one about
moments in section \ref{s:Iso}, which is not surprising since moments implement
the canonical analog of higher time derivatives.)

The main question regarding higher spatial derivatives is not whether they
would undo the higher-order deformation, but whether they can be consistently
implemented at all. Attempts to do so in spherical symmetry
\cite{HigherSpatial}, where such expansions have been formulated
systematically, indicate that anomaly-free versions with higher spatial
derivatives are much more complicated to achieve than higher-order versions,
but of course, they may not be impossible.\footnote{Alternatively, one
  could try to find a consistent deformed version of a discrete
  hypersurface-deformation algebra, building on \cite{DiscDirac}. However,
  even if such a system could be found with loop modifications, it would not
  give rise to a continuum effective theory.}  If no consistent versions with
higher spatial derivatives existed, loop quantum gravity and its premise of
non-local basic holonomy operators would be ruled out.

\subsubsection{Evaluation of equations}

As with any first-class constrained system, there are different options to
evaluate a consistent deformation of the classical hypersurface-deformation
algebra for physical consequences: (i) One can try to find gauge-invariant
variables and solve their equations of motion. (ii) One may pick a gauge of
the consistent gauge system, or select an internal time and deparametrize in
order to compute relational observables. The second option is usually easier
to implement, but also the first one is feasible for linear
perturbations. Both procedures rely on the consistency of the constraint
algebra, for if there were anomalies (or if one just inserted modifications
without checking for anomalies) gauge-invariant variables would not exist, and
results would be unphysical because they would depend on which gauge or
internal time one had selected.

Gauge fixing and deparametrization cannot be used reliably before a
consistent representation or deformation of the algebra has been found, for
this would prevent one from making sure that there are no anomalies. (Partial
gauge fixings may be used to find hints at potential forms of consistent
deformations with reduced effort \cite{ScalarHolEv}. However, there are no
intrinsic means to check consistency.)  It is unlikely that any consistent
system results if one uses classical gauge transformations to set up the
system but then modifies the dynamics, which for generally covariant theories
is part of the gauge content. In this respect, quantum gravity differs
crucially from other gauge theories in which (Gauss-like) constraints take much
simpler forms unrelated to the dynamics. If gauge fixing or deparametrization
are used carefully before inserting quantum modifications, a formally
consistent system might be obtained. But one can ensure anomaly-freedom only
by showing explicitly that results do not depend on the chosen gauge or
internal time, and such a procedure is usually not feasible for the
complicated systems subject to hypersurface deformations. Current models that
do use gauge fixing or deparametrization in this way do not ask what would
happen if one had made a different choice.

There is another subtlety tightly related to closure of the constraint algebra
and the existence of gauge-invariant variables. Hamilton's equations
explicitly contain the lapse function and shift vector. Since neither of the
two is a phase-space variable, their gauge transformations are not readily
available. On the other hand, different choices of the lapse and shift result
in different definitions of both the geometrical and matter canonical
variables.  It is thus clear that the lapse and shift should be gauge
transformed, too, for equations of motion to be consistent.  In fact, one can
deduce such transformations by changing the gauge of Hamilton's
equations. Since on the left-hand side there is always a time derivative
generated by a constraint, a change of gauge of the entire equation depends on
the constraint algebra. As shown, for example, in \cite{ScalarGaugeInv}, there
are usually (many) more canonical equations than the number of lapse and shift
components. Therefore, it is quite non-trivial to make all the equations of
motion transform in a consistent way for one fixed lapse and shift
transformation. Remarkably, this is possible owing to the closure of the
constraint algebra. (See also \cite{LapseGauge} for an alternative approach to
this issue, treating the lapse function and shift vector as additional
canonical variables, thus expanding the phase space.)

First-class constraints, therefore, cannot be solved or otherwise eliminated
before one implements modifications. Second-class constraints, on the other
hand, play a different role, just as they do in the context of
quantization. Such constraints are relevant when one uses reduced models to
expand around or to compute consistent algebras in, such as spherically
symmetric systems. The requirement that non-symmetric modes and their momenta
vanish then contributes second-class constraints to the original first-class
system. (It is sometimes stated that symmetric models follow from partial
fixings of the diffeomorphism gauge. This is not true, for gauge fixing would
not remove physical degrees of freedom, as must happen for a reduced model to
be defined.) In contrast to first-class constraints, second-class constraints
are most easily solved before quantization or modification. They do not
generate gauge transformations, and therefore the issue of gauge dependence
clearly does not appear. Moreover, when they are used for reduced systems,
they do not imply anything about space-time structure. They just remove
degrees of freedom but do not change the structure of remaining
ones. Midi-superspace models then still allow one to explore quantum
space-time.

\subsubsection{Space-time}

At present, it is not clear whether there is a generalized space-time picture,
such as a non-commutative \cite{Connes,NonCommST} or a fractional one
\cite{Fractional,FractCosmo}, suitable for the deformation (\ref{HHbeta}). In
some regimes, as shown in \cite{DeformedRel}, the deformation produces effects
comparable to doubly special relativity \cite{DSR1,DSR2,DSR} with a deformed
Poincar\'e algebra in the sense of \cite{GeneralizedPoincare}. This shortage
of space-time models does not pose problems for physical evaluations of the
theory, which can all be done at the canonical level once an anomaly-free set
of constraints is given. Nevertheless, a space-time model would be useful in
that it could provide additional intuition.

It is clear that any deformation in (\ref{HHbeta}) would imply that standard
space-time diffeomorphisms or coordinate changes are no longer gauge
transformations of the effective metric: The algebra fails to correspond
to Lie derivatives along space-time vector fields, even on the constraint
surface. (It might, however, be possible to represent gauge
  transformations as Lie derivatives of redefined variables, as realized for
  instance in the examples of \cite{Absorb}.) This fact implies that the
underlying structure cannot be Riemannian, and indeed, any attempt to write a
line element
\begin{equation} \label{LineEl}
 {\rm d}s^2=g_{ab}{\rm d}x^a{\rm d}x^b
\end{equation}
with $g_{ab}$ subject to modified gauge transformations of the deformed
algebra would fail to produce an invariant. One would have to modify
coordinate transformations in such a way that they, together with the modified
gauge transformations of $g_{ab}$, cancel in the combination used for ${\rm
  d}s^2$. We emphasize that, in order to capture the quantum effects under
  discussion, it is not sufficient to allow for $g_{ab}$ to satisfy a
  quantum-corrected version of Einstein's equations. The classical metric
  tensor or curvature quantities computed from it are no longer covariant when
  the classical gauge algebra is deformed. (For this reason, curvature
  singularities in classical models of signature change, as recently
  considered in \cite{SigChangeHybrid}, do not have much to say about the
  regularity of our models of signature change with deformed space-time
  structures.) In the absence of a suitable space-time tensor calculus based
  on line elements (\ref{LineEl}), invariant quantities would have to be
  computed by canonical means, which can easily be done at the level of
  cosmological perturbations following \cite{ScalarGaugeInv}.

Although the conclusions of \cite{Regained,LagrangianRegained} no longer apply
with a deformed algebra (\ref{HHbeta}), the methods can still be used
\cite{Action}. In this way, one can reconstruct a Lagrangian density ${\cal
  L}$ corresponding to the algebra, but an action is more difficult to obtain
because the integration of ${\cal L}$ requires one to know the deformed
space-time structure and a description in terms of coordinates or covariant
measures. Even so, such reconstructed Lagrangian densities provide further
insights. For a scalar field, for instance, covariance under the deformed
algebra (\ref{HHbeta}) implies that a second-order Lagrangian density can
depend on derivatives of the field only via the combination
\begin{equation} \label{Deriv}
 \beta\chi-(\partial_{\perp}\phi)^2
\end{equation}
with the normal derivative $\partial_{\perp}\phi=N^{-1}\{\phi,C[N]\}$ and the
spatial derivatives $\chi:=h^{ab}(\partial_a\phi)$ $\times(\partial_b\phi)$. 
  (The derivation in \cite{LagrangianRegained,Action} requires that $\beta$
  does not depend on the momentum of the metric. Deriving a generic derivative
  term for mode equations in the presence of holonomy corrections is therefore
  more complicated.) This modified derivative term makes it clear that the
theory has a non-classical notion of covariance, but one that is fully
consistent thanks to the anomaly-free deformation of the algebra.

The same modification appears in the gravitational mode equations, which can
be derived from a specific consistent version of the modified constraints. For
pure holonomy corrections in (\ref{HQcosmo1})--(\ref{HQcosmo2}), for instance,
we have $\beta(\bar{k})= \cos(2\delta \bar{k})$ in the deformed algebra. One
can then derive the correction to the Mukhanov--Sasaki equation of motion for
gauge-invariant perturbations of scalar and tensor type $v_{\mathrm{S,T}}$
\cite{ScalarTensorHol}. (These perturbations are invariant under the deformed
gauge transformations generated by modified constraints. For such invariant
variables to exist, off-shell closure of the constraint algebra is essential.)
In conformal time $\eta$, this is given by
\begin{equation}\label{EoM}
{v}''_{\mathrm{S,T}} - \beta \, \nabla ^2 v_{\mathrm{S,T}} -
\frac{{z''_{\mathrm{S,T}}}}{z_{\mathrm{S,T}}} v_{\mathrm{S,T}}= 0 \,, 
\end{equation}
which reduces to the classical equation when $\beta\to1$. Here, primes are
derivatives with respect to conformal time. This equation holds for both
scalar and tensor perturbations, with suitable background functions
$z_{\mathrm{S,T}}$.  (In the presence of inverse-triad corrections, scalar and
tensor modes are subject to wave equations with different speeds
\cite{LoopMuk,ScalarHolInv}.)  For scalar perturbations, the Mukhanov
variables in the deformed case are given by
\begin{equation}
v_{\mathrm{S}}=\sqrt{\bar{p}}\,\left(\delta\phi +
  \frac{{\bar\varphi}'}{{\cal H}}\phi\right) 
\quad \mbox{and}\quad
 z_{\mathrm{S}} \;=  
\; \sqrt{\bar{p}}\, \frac{{\bar{\varphi}}'}{{\cal H}}
\end{equation}
where ${\cal H}$ is the Hubble parameter in conformal time.  For tensor modes,
one obtains
\begin{equation}\label{vzt}
v_{\mathrm{T}}=\sqrt{\frac{\bar{p}}{|\beta|}} h
\quad \mbox{and}\quad
 z_{\mathrm{T}} \;= 
\; \sqrt{\frac{\bar{p}}{|\beta|}}\,,
\end{equation}
where $h$ represents the two degrees of freedom.  Inserting (\ref{vzt}) into
(\ref{EoM}), we obtain the following equation of motion for tensor
perturbations:
\begin{equation} 
(h^i_a)'' + (h^i_a)'
 \; \left(2{\cal H} - {\frac{\beta'}{\beta}}\right) - \beta \, \nabla ^2
 {h^i_a} = 0\,. 
\end{equation}

The modified space-time picture, although not directly observable, has further
physical consequences. Most importantly, since time translations are
intimately related to energy conservation, the status of this important law is
unclear when constraints of the hypersurface-deformation algebra are
modified. If the algebra remains consistent, energy conservation is still
realized in the general sense of continuity equations \cite{Energy}, but
without considering the algebra no such statement could be made.

\subsubsection{Signature change}
\label{s:Sig}

The derivative term (\ref{Deriv}) in a scalar Lagrangian density, as well as
explicit derivations \cite{ScalarHol,ScalarTensorHol} of equations for tensor
modes (\ref{EoM}), shows that covariance and the form of dispersion relations
are modified in a theory subject to (\ref{HHbeta}). There is an even more
dramatic consequence, owing to the fact that $\beta$ can become negative at
high density in the presence of holonomy modifications
\cite{JR,ScalarHol}. Equations of motion for modes then become elliptic rather
than hyperbolic differential equations, and no well-posed initial-value
problem exists. In particular, in anomaly-free cosmological models one cannot
evolve through high-density regimes such as bounces. In terms of space-time,
this change of signature can be seen from the fact that a relation
(\ref{HHbeta}) with $\beta=-1$ is obtained for 4-dimensional Euclidean
space. (But note that there is no piece of {\em classical} Euclidean space if
$\beta=-1$ is found only on one maximum-curvature slice. One should therefore
not expect an effective Hamiltonian constraint to resemble the classical
Euclidean one.
Moreover, the space-time structure is strongly non-classical near the
transition where $\beta=0$. If $\beta$ changes smoothly through zero, no
singularities arise in the canonical background and mode equations, in
contrast to classical models of signature change as in \cite{SigChangeClass}
or in the recent \cite{SigChangeHybrid}.)

\paragraph{Origin:}

Signature change appears to be a strong consequence of holonomy modifications,
which had been overlooked until spherically symmetric inhomogeneity and
cosmological perturbations were studied in an anomaly-free way. Indeed,
without inhomogeneity, one cannot determine the signature because (i) one
cannot see the relative sign between temporal and spatial derivatives and (ii)
the relation (\ref{HH}) trivially equals zero in homogeneous
models. Nevertheless, signature change is not a consequence of inhomogeneity,
the latter rather being used as a test field. One can see this easily from the
derivative term (\ref{Deriv}) or the one in (\ref{EoM}) for negative
$\beta$, it remains positive definite no matter how small the spatial
derivatives are (unless they are exactly zero). Signature change is a
consequence of the strong modification one makes if one uses holonomy terms at
high density, not of a small amount of perturbative inhomogeneity added to the
system.

To bring out this latter point more clearly, one may compute the magnitude of
modifications that holonomy terms imply compared to the classical Friedmann
equation. Instead of $H^2=(8\pi G/3) \rho$ for the classical
squared Hubble parameter $H=\dot a/a$, in the presence of holonomy
modifications we have
\begin{equation} \label{FriedHol}
\frac{\sin^2(\ell K)}{\ell^2}= \frac{8\pi G}{3}\rho\,,
\end{equation}
in which the sine function and the length parameter $\ell$ (which may be
related to the Planck length and also includes the Barbero--Immirzi parameter)
are subject to ambiguities. The argument $K$ of the sine function is related
to the momentum canonically conjugate to the scale factor, and can be viewed
as a modified version of the Hubble parameter which equals $H$ for $\ell\to0$.
We can therefore view both the classical equation and the modified version as
resulting from functions of $K$, the forms of which are given by different
versions of canonical constraints: $C_{\rm class}(K)=K^2$ and $C_{\rm
  mod}(K)=\ell^{-2}\sin^2(\ell K)$. In order to provide a quantitative
comparison of these functions, we compute the ratio $(C_{\rm class}(K)-C_{\rm
  mod}(K))/C_{\rm class}(K)$ for some characteristic values of $K$.

Maximum density is reached when the sine reaches the value $\sin(\ell K)=1$,
so that $\ell K=\pi/2$, giving for $\ell=\ell_{\rm P}$ the maximum density
$\rho_{\rm max}=3\rho_{\rm P}/(8\pi)$. At this density, the relative magnitude
of the modification, replacing $K^2$ by $\sin^2(\ell K)/\ell^2$, is
\begin{equation}
 \frac{K^2-\sin^2(\ell K)/\ell^2}{K^2}=
 1-\frac{\sin^2(\ell K)}{\ell^2K^2}=
 1-\left(\frac{2}{\pi}\right)^2 \approx 0.6\,.
\end{equation}
The signature changes when $\cos(2\ell K)=0$, or $\ell K=\pi/4$, implying a
relative magnitude of the modification given by $1-\sin^2(\ell K)/(\ell K)^2 =
1-8/\pi^2\approx 0.2$. In the Euclidean regime, holonomy modifications in the
Friedmann equation change the classical term by a contribution between $20\%$
and $60\%$, which is certainly much larger than any inhomogeneity one would
add in the perturbative regime. It is this background modification that, when
embedded in a consistent and anomaly-free theory of inhomogeneity, implies
drastic changes to the structure of space-time. As long as the background
modification stays below $20\%$, space-time is Lorentzian no matter how large
the inhomogeneity is (within the perturbative regime). If the background
modification exceeds $20\%$, the signature changes even for tiny
inhomogeneity.

\paragraph{Instability:}

A change of sign in front of the Laplacian of the Mukhanov--Sasaki
  equations is not a novelty in cosmology. The same effect occurs, for
  instance, in higher-order gravity where the Gauss--Bonnet curvature
  invariant is non-minimally coupled with a scalar field (\cite{CdD} and
  references therein). In that case, this change of sign is simply interpreted
  as a classical instability of the perturbations on the FRW background
  affecting cosmological space-time scales. Within the same model, one can
  have further modifications to the Mukhanov--Sasaki equations which may
  introduce ghost and tachyon instabilities as well as superluminal
  propagation. Although all of these features are problematic and can be
  avoided by a restriction of the parameter space, the nature of the
  space-time wherein perturbation modes propagate remains purely classical and
  Lorentzian. (The theory is of higher-curvature type and does not lead to
  deformations of the constraint algebra.)

  Here, however, the change in the perturbation equations is a direct
  consequence of the deformation of the constraint algebra of gravity and,
  hence, of a deformation of the classical space-time structure. The type of
  field equations changes from hyperbolic to elliptic for all modes
  simultaneously and does not depend on the physical meaning of a
  mode. Moreover, the manifold on which physical fields are defined no longer
  allows any causal structure at high curvature. This is the main reason why,
  in the present context, such an effect is regarded as a signature change of
  space-time rather than a simple perturbation instability.

  In practical terms, signature change implies instabilities because the
  oscillating time dependence $\exp(\pm i\omega t)$ is replaced by an exponential
  $\exp(\pm\omega t)$. The growing mode leads to an instability if
  initial values are chosen at some fixed $t$. (A boundary-value problem
  fixing the field at two different values of $t$, as suitable for elliptic
  equations, eliminates the instability.)  The significance of the instability
  depends on the $t$-range during which the growing mode is realized. If the
  $t$-range is of the order of a Planck time, for instance, only
  trans-Planckian modes will be affected by the instability and the problem
  may appear less severe.\footnote{This question goes back to a suggestion by
    Jaume Garriga.} (The stability issue would be a version of the
  trans-Planckian problem.)

  One can estimate the $t$-range in the solvable background model of
  \cite{BouncePert}, in which a free massless scalar $\phi$ with momentum
  $p_{\phi}$ in a holonomy-modified spatially flat isotropic model is shown to
  imply a volume expectation value $\langle\hat{V}\rangle(\phi)\propto
  \cosh(\phi-\phi_0)$ with a constant $\phi_0$. This background dynamics
  implies signature change of inhomogeneous modes when the free scalar density
  $\rho_{\rm free}=\frac{1}{2}p_{\phi}^2/V^2$ is between one half and the full
  Planck density; accordingly, $\langle\hat{V}\rangle$ changes by a factor of
  order one. (The momentum $p_{\phi}$ is constant for a free massless scalar.)
  Therefore, $\phi$ changes by adding a constant of order one. Using the
  equation of motion ${\rm d}\phi/{\rm d}t=p_{\phi}/V=\sqrt{2\rho_{\rm
      free}}\sim \sqrt{\rho_{\rm P}}$ in the high-curvature phase, the
  $\phi$-range of order one translates into a $t$-range of about a Planck
  time. Only trans-Planckian modes are then affected. However, if one tries to
  extend these arguments qualitatively to more general backgrounds
  \cite{FluctEn}, one can see that the general $t$-range is larger. The
  relationship between $\phi$ and $t$ is still given by the previous equation,
  where now ${\rm d}\phi/{\rm d}t=p_{\phi}/V=\sqrt{2\rho_{\rm kin}}$ is
  related to the kinetic energy density of the field. If there is a potential,
  or even just quantum fluctuations which one should expect in the
  high-curvature regime, the kinetic density would only be a small part of the
  total density and the $t$-range would be correspondingly larger for the same
  $\phi$-range. More and more modes of sub-Planckian frequencies will be
  affected by the instability, a problem which can be solved only by using a
  boundary-value problem as suitable for an elliptic differential equation.

We would like to stress that, due to the incomplete status of the theory, at present the signature change remains a possibility.
It may turn out that there is a mechanism within loop quantum gravity by which signature change could be avoided after all.\footnote{Even if this were the case, one could not show it within mini-superspace models because one must have access to temporal and spatial variations. This is the reason why mini-superspace models of loop quantum cosmology (not necessarily Wheeler--DeWitt) do not seem to be completely viable at high density.} At present, however, this seems unlikely because the results, not just from the effective approach but also from operator computations of off-shell constraint algebras, appear to be very universal and generic.

\subsubsection{Phase space and the Hojman--Kucha\v{r}--Teitelboim theorem}

It may be useful to stress that in the effective-constraints approach one does not impose the same phase-space degrees of freedom as in classical general relativity. The only important assumption made concerning this point is that loop quantum gravity allows for a sufficiently large class of semi-classical states in which the basic operators of the quantum theory have certain expectation values with small fluctuations. If this assumption was violated, the theory could not be considered as a consistent version of quantum gravity. If the assumption is satisfied and LQG therefore has a chance of being consistent, the expectation values just mentioned serve as variables which allow us to formulate a phase space for an effective theory. In a perturbative treatment as described in sections \ref{sss:Cosmpert} and \ref{s:Pert}, the phase space we find is truncated due to the way perturbations are handled. To the $\hbar$-orders considered here, this phase space simply happens to coincide with the classical one. 

Even if the effective phase space of the present approach and the one of general relativity match to lowest $\hbar$ order, the Hojman--Kucha\v{r}--Teitelboim theorem of \cite{Regained} does not apply. The two main assumptions of \cite{Regained} are that (a) the degrees of freedom are the same as in general relativity and that (b) the space-time structure (in the sense of hypersurface deformations) is \emph{classical}. Then, the Hojman--Kucha\v{r}--Teitelboim theorem shows that the classical dynamics for second-order field equations is unique and given by Einstein's theory. When one considers canonical quantum gravity or its effective theories, the hypothesis that space-time remains classical is, of course, questionable. If assumption (b) is dropped, no uniqueness of constraint algebras can be achieved. This is the context in which we discuss more general versions.

\subsubsection{Possible observations}

There are two main paths to probe possible observational consequences of
quantum gravity in general, and also of loop quantum gravity or cosmology. One
is to rely on ``low energy'' effects but somehow compensate for their
minuscule nature by looking at phenomena with a large relevant parameter of
classical form, for instance huge times or distances over which those tiny
effects, when integrated, can lead to measurable consequences. This idea is
the basis for searches of Lorentz-invariance violations; for a review on such
attempts in loop quantum gravity, see \cite{LQGPheno}.  However, as neither a
clear proof that Lorentz-invariance violations are indeed expected in loop
quantum gravity, nor experimental evidence of such an energy-dependent time
delay were established, we rather focus on the other main path: observational
cosmology.

Cosmology involves a long time scale and is therefore partially related to the
first type of phenomenology just described. However, at the same time it has
the potential of allowing access to more energetic phenomena in the very early
universe. Loop quantum cosmology indeed suggests strong modifications to the
classical dynamics of space-time, as shown for instance by popular bounce
models.  To go beyond these background effects, perturbations by inhomogeneity
have to be calculated. The framework reviewed in this article is compatible
with a rigorous treatment of perturbations. Once the equations of motion for
perturbations are known and a reasonable assumption for the vacuum state is
made (section \ref{s:InflVac}), the primordial power spectrum can be
calculated. The latter can then be used as an input to evaluate the subsequent
multipole spectrum $C_l$ which can directly be compared with
observations. Then, either a MCMC or a Fisher analysis can be performed to
compare the prediction with existing or forthcoming data.

As described earlier, there are two main classes of effective corrections that
have been studied so far in loop quantum cosmology: holonomy and inverse-triad
corrections (see \cite{SIGMA,GCPheno,ObsLQC} for reviews). As usual, there are
three kinds of perturbations (scalar, vector, tensor) which decouple at the
first order. Of course, other corrections, especially quantum back-reaction,
might play important roles. Many articles have been devoted to computing
primordial spectra in this framework of deformed algebras. Most (but not all)
of them deal with tensor modes
\cite{BounceTensor,TensorHalfII,BounceTensorII,GravWaveLQC,Observing}. In
particular, it was shown in \cite{Omega} that when the more-restricted
deformation due to scalar modes is taken into account for tensor modes, the
situation drastically changes. By regularizing the equation of motion, the
resulting primordial spectrum was shown to be scale invariant in the IR limit,
exponentially rising in the UV limit and oscillating in between. These effects
are mainly due to high curvature.

Inverse-triad corrections are not directly related to curvature and could
potentially be large in more-accessible regimes. From this perspective, they
have been investigated in \cite{LoopMuk,InflTest,InflConsist}.

Deformed general relativity not only implies modified gravitational dynamics
but also a non-classical space-time structure. In addition to effects in
equations of motion, there may be traces of quantum gravity in the choice of
initial states for inflation. After all, the common selection of a vacuum
state in the infinite past relies on quantum field theory in curved but
classical space-time. With deformed space-time structures, several features
usually taken for granted in quantum field theory change, for instance those
of stress-energy tensors \cite{Energy}. The treatment of states (initial or
otherwise) must therefore be examined critically, as we will do in
section \ref{s:States}.

\subsection{Add-on inhomogeneity}
\label{s:Add}

The derivation of an anomaly-free algebra is complicated and may be impossible
in full generality. Several shortcuts to cosmological perturbations with loop
modifications have therefore been pursued. We have already commented on the
most obvious one (gauge fixing before implementing modifications) and
its failure to capture a consistent space-time structure. The two other
approaches discussed in this subsection are more subtle, but they both rely on
deparametrization without checking for independence on one's choice of
internal time.\footnote{The ambiguities regarding choices of constrained systems with the same constraint surface may seem, at first sight, to pose a similar problem to the choice of internal time in deparametrized models. Different constrained systems would be regarded as different theories which happen to have the same classical field equations. The choice of constrained system therefore amounts to a quantization ambiguity. Different choices of internal times partially correspond to quantization ambiguities too, but here also a lack of covariance (for some choices) plays a role. One can see this in homogeneous models, which have just one constraint and therefore a simple constraint hypersurface. Yet, different choices of internal times can easily lead to inequivalent quantizations. Our main point is that deparametrization does not make it easy to disentangle different kinds of ambiguities, while off-shell constraint algebras (classical, effective or quantum) enjoy a wider set of established techniques.}

\subsubsection{Hybrid models}\label{hyb}

Hybrid models \cite{Hybrid} combine a Fock quantization for (non-perturbative)
inhomogeneity with a loop-quantized and therefore modified homogeneous
background. Gauge fixing is used to obtain a clear separation between
background and inhomogeneity.  This avenue has been evaluated in quite some
detail in Gowdy models \cite{Gowdy,GowdySing,GowdyQuadratic}, for which
several results on standard quantizations and their unitarity already exist
\cite{UniqueSchroedGowdy,UnitaryGowdyFeas,UniqueFockNonStat}. For loop
methods applied to Gowdy models, see
\cite{EinsteinRosenAsh,EinsteinRosenQuant}.

A hybrid approach amounts to quantum field theory on a modified background,
which, unlike loop-quantized inhomogeneity, does not lead to problems with
space-time covariance and anomalies. After all, covariant quantum field theory
can be formulated on any curved background \cite{LocalQuant} without requiring
Einstein's equations to hold. From the point of view of space-time structure,
the Fock-quantized inhomogeneity of hybrid approaches plays the role of a
matter field on some Riemannian space-time. Indeed, in spite of the use of
gauge fixing, hybrid approaches are formally consistent, and it is even
possible to compare several choices of gauge fixing and show that the
resulting quantum theories are unitarily related \cite{HybridFlat}.

While hybrid models provide consistent treatments of gauge systems, from the
point of view of loop quantum gravity the truncation used appears to be too
restrictive to study evolution of inhomogeneity through high density.  When
holonomy modifications are relevant for the background, for instance near a
potential bounce, they should be important in mode equations as
well. Fock-quantized inhomogeneity ignores the latter corrections, and
therefore one can view hybrid models as models for loop quantum gravity only
when modifications in the background are weak, restricting their use to
low-curvature regimes in which the dynamics does not differ much from standard
quantum field theory. Nevertheless, one can formally extend the evolution to
high density, and this brings out one encouraging feature: Inhomogeneity
appears to evolve through high density in an unstable manner
\cite{InhomThroughBounce}, just as one expects when an evolution or
initial-value picture is applied in a non-hyperbolic regime (see
section \ref{s:Sig}). Although the equations of hybrid models do not directly
show signature change, their solutions seem to give rise to somewhat similar
effects.

\subsubsection{Dressed metric}

The authors of \cite{AAN} attempt to incorporate loop modifications in
evolution equations for perturbative inhomogeneity, and they are careful in
avoiding gauge fixing (but still use deparametrization).  This approach is
intended to deal consistently with quantum fields on a quantum background,
building upon \cite{QFTCosmo} (see also \cite{QFTCosmoClass}). The result is a
possible method to derive cosmological predictions in a ``less effective'' way
that uses full wave functions. However, one can wonder if there is a
consistent space-time structure in their framework as, instead of fixing the
gauge, the authors solve classical constraints for gauge-invariant modes
before they implement loop modifications and quantize the deparametrized
Hamiltonian. The difference between classical gravity (even when linearized) and fields on a fixed background is that in the former setting one specifies how the fields transform under (linear) coordinate transformations. The same holds at the quantum level, and a consistent space-time structure is what sets apart a quantum-gravity model from a version of quantum-field theory on a possibly modified background. In deparametrized models, the evolution
generator is a combination of some terms in the constraints. If the former is
modified, it is questionable to assume that the constraints and perturbative
observables are still classical.\footnote{As with gauge fixing, it is
  sometimes legitimate to solve classical constraints before quantization. For
  instance, at the classical level, the Gauss constraint may be solved
  explicitly to rewrite the spin connection in terms of the densitized
  triad. Importantly, this possibility remains intact at the effective level
  for the following reason: In the considered quantization schemes, the Gauss
  constraint does not acquire any quantum corrections, thus the expression for
  the spin connection is identical to its classical counterpart.} The
resulting equations may be formally consistent, but just as with gauge-fixed
modifications, a corresponding space-time picture might not exist. Not so
surprisingly, the authors see no hint of signature change and pose an initial
state at the bounce point of background evolution, where the initial-value
problem is hard to define if the inhomogeneities effectively live in a quantum
version of four-dimensional Euclidean space at this point.

Although shown to be mathematically equivalent to some quantum field theory on
a quantum geometry (a classical-type metric with quantum-corrected
coefficients), the ``dressed metric'' approach \cite{AAN} assumes, in some
sense implicitly, a standard space-time structure. The validity of this
assumption rests on an undeformed constraint algebra, whose very existence can
be questioned in the presence of the holonomy modifications used: As shown by
the derivation of anomaly-free modified constraints, especially in
\cite{ScalarHol,ScalarHolInv}, the algebra of constraints is generically
deformed by quantum-geometry corrections. Current derivations are not
complete, as described before in this article, but the mere possibility of
deformations shows that standard space-time structures cannot just be assumed
but would have to be derived. This is not done in the approach of
  \cite{AAN}.

The authors of \cite{AAN} criticize derivations of anomaly-free constraint
algebras, and in fact the whole underlying approach. Most of this criticism
had already been addressed in the literature, although in somewhat dispersed
locations.\footnote{The claim made in \cite{AAN} that the anomaly-free algebra
  of \cite{ConstraintAlgebra} is not deformed is in fact not correct as the
  algebra is deformed as in (\ref{HHbeta}), albeit with a positive $\beta$
  that does not give rise to signature change.} Our earlier discussions in the
present article summarize and strengthen previous statements and respond
to the remaining issues raised in \cite{AAN}, making the case why the result
of deformed algebras should be taken seriously. (We remind that the
results of \cite{Regained} do not underline any inconsistency in the deformed-algebra approach, as both the constraints and
the algebra are simultaneously deformed in a consistent way, as discussed previously.) Moreover, although higher-time derivatives have not yet been included
in current deformed versions, owing to the increased complexity of
calculations, there is a clear procedure to do so within the general scheme of
canonical effective equations. (The approach of \cite{AAN} does not include
higher time derivatives either, and it does not specify a clear method for
doing so.)

The incompleteness of present anomaly-free constraint algebras and the
specific way of introducing counterterms may certainly be criticized. (See
section \ref{s:Examples} for a discussion of the latter.) But the issue of
anomaly freedom cannot simply be ignored and should in some way be faced also
in approaches such as \cite{AAN}. A fully consistent treatment of loop quantum
gravity and its cosmological applications is clearly still missing.  The
progress made recently would benefit from a global understanding of the
different issues and subtleties. No proposal made at this time addresses all
of them.

\subsubsection{Relation to possible observations}

 It is conceivable that observing the anisotropies of the B-mode cosmic
  microwave background may help to discriminate between the different
  theoretical approaches to loop quantum cosmology previously mentioned. One
  can compute the primordial power spectra for the ``dressed metric'' approach
  and frameworks which lead to instabilities (such as the hybrid approach or
  signature-change models in which one uses an initial-value problem
  throughout the elliptic phase; in the following, we consider this class of proposals as one approach). So far, this has been explicitly done
  assuming holonomy corrections only. For a proper comparison, the initial
  state of the inhomogeneous degrees of freedom can be chosen as the Minkowski
  vacuum in the infinite past of the contracting phase, since this allows one to
  set the {\it same} initial conditions for both approaches. (This would
  not be possible if the initial state was selected at the time of the bounce,
  for the semi-classical structure of the background space-time in deformed
  general relativity drastically differs from the semi-classical structure of
  the background used in the dressed-metric approach.) For such a choice of
  the initial conditions, existing calculations suggest that the predicted
  power spectra for tensor modes share common features but, more importantly,
  also exhibit possibly drastic differences, depending on the scales. For
  wavenumbers $k<\sqrt{8\pi G\rho_{\rm max}}$ ({\it i.e.} large scale),
  both approaches for treating cosmological perturbations lead to a nearly
  scale-invariant (slightly blue-tilted) spectrum followed by
  oscillations. However, for smaller scales $k>\sqrt{8\pi G\rho_{\rm max}}$,
  the deformed general relativity approach yields an exponential increase
  while the standard, slightly red-tilted inflationary power spectrum is
  recovered in the dressed-metric framework. (It should be noted that this UV
  behaviour of the primordial power spectrum in the dressed-metric approach is
  also obtained by fixing the initial state at the time of the bounce
  \cite{AAN}.) Such a difference could be clearly observed in the angular
  power spectrum of the polarized CMB anisotropies, showing that, at least in
  principle, cosmological observation may be used to potentially discriminate
  between the considered approaches. 

  It is less clear how one could distinguish results of hybrid models from
  those implied by deformed constraint algebras, given that the former seem to
  give rise to the same kind of instability. However, the cosmology of
  deformed constraint algebras still has to be fully analyzed, making use of a
  well-posed initial/boundary-value problem of mixed type. Some part of an
  initial state in the infinite past affects the expansion phase, but so do
  boundary values chosen for the elliptic regime. Without a detailed analysis,
  it is difficult to say if such a setup would make instabilities noticeable
  in power spectra.

\section{States}
\label{s:States}

Canonical effective equations describe the evolution of a quantum state via
the time dependence of expectation values of basic operators and their moments
instead of wave functions. Correction terms compared to classical equations
then depend via the moments on the state used, at least implicitly. A
completely general state is difficult to parametrize, and therefore
additional assumptions are often made to restrict possible choices. The most
well-known one is the choice of a near-Gaussian state in low-energy effective
actions, motivated physically by the nature of free vacuum states. In quantum
cosmology, the question of states appears in two different forms, regarding
the initial state of an inflaton and the quantum-gravity state of an evolving
space-time background.

\subsection{Inflaton vacuum}
\label{s:InflVac}

A standard ingredient of the inflationary scenario is the selection of initial
conditions according to vacuum fluctuations of the curvature
perturbation. These initial conditions may directly be combined with effective
equations which describe the evolution of a state. Nevertheless, it is
sometimes seen as awkward that a treatment of effective descriptions, working
with quantum corrections to equations of classical type, refers to a wave
function or full quantum state in some of its parts. One occasionally, and
quite misleadingly, describes such a procedure as a ``re-quantization'' of an
effective theory, which is supposed to be an approximation to a theory that
was quantum to begin with. The separation between initial conditions and
equations makes it clear that there is no such quantization after selecting
initial conditions corresponding to some quantum state, and therefore the
usual treatment in inflationary models does not pose a problem. Nevertheless,
it is instructive to discuss how vacuum initial values can be derived at the
level of effective equations.

At this stage, the feature of canonical effective equations describing the
evolution of a quantum state characterized by its moments, as recalled in
section \ref{s:Degrees}, becomes important. Not only are expectation values
subject to equations of motion correcting the classical dynamics, but moments
of an evolving state have a dynamics of their own. Moreover, unlike the
expectation values or classical variables, they are subject to quantum laws
such as the uncertainty relation
\begin{equation}
\Delta(q^2)\Delta(p^2)-\Delta(qp)^2\geq \frac{\hbar^2}{4}
\end{equation}
for a canonical pair.  This relation, formulated directly for moments as they
feature in effective equations, together with the dynamical statement of
short-term stationarity of any ground state, allows one to select initial
conditions for effective equations (see also \cite{CUP,Wiley}).

We briefly recollect the relevant classical equations. Fourier modes of the
curvature perturbation $v$ evolve according to the Mukhanov--Sasaki equation
\begin{equation} \label{MS}
 v_k''+\left(k^2-\frac{z''}{z}\right)v_k=0,
\end{equation}
where $z$ is related to the scale factor $a$. At large scales, ignoring the
$k^2$-term, the solution must take the form $v_k\sim A(k)z$ with a
normalization function $A(k)$ to be determined. One usually treats $v_k$ like
a wave function and imposes conditions that amount to the Bunch--Davies vacuum
in the infinite past, implying $A(k)= \exp(-ik\eta)/\sqrt{2k}$. The power
spectrum depends on $|v_k|^2=[z/z(k_*)]^2/(2k)$, where $z(k_*)$ is taken at
Hubble crossing.  The normalization of $v_k$ is motivated by the inflationary
postulate that $v$ initially have a magnitude given by vacuum fluctuations.

The same postulate allows us to derive the normalization condition from
effective equations, not making use of wave functions.  For very early times
($\eta\to-\infty$), Eq.\ (\ref{MS}) is generated by a harmonic-oscillator
Hamiltonian with ``frequency'' $k$ and ``mass'' $m=1$. Ground-state
fluctuations then take the value $|\Delta v_k|=\sqrt{\hbar/(2k)}$, which is
taken as the value of $A(k)$. It is now important to note that one need not
re-quantize the mentioned Hamiltonian in order to derive $\Delta v$. Canonical
effective equations for the harmonic oscillator imply
\begin{eqnarray}
\frac{{\rm d}\Delta (q^2)}{{\rm d}t} &=& \frac{2}{m}\Delta(qp)\,,\\
\frac{{\rm d}\Delta(qp)}{{\rm d}t} &=&
\frac{1}{m}\Delta(p^2)-m\omega^2\Delta(q^2)\,,\\
\frac{{\rm d}\Delta(p^2)}{{\rm d}t} &=& -4m\omega^2\Delta(qp)\,.
\end{eqnarray}
For a stationary state, all these time derivatives vanish, and therefore
$\Delta(qp)=0$ and $\Delta(p^2)=m^2\omega^2\Delta(q^2)$. Together with the
requirement that the uncertainty relation be saturated for the harmonic
(free-theory) ground state, we then have $\Delta(q^2)= \hbar/(2m\omega)$,
which amounts to $|\Delta v_k|^2=\hbar/(2k)$ with the values of ``mass'' and
``frequency'' quoted above.

We used only effective equations to derive the magnitude of quantum
fluctuations, applied to the harmonic oscillator to model the free
vacuum of field modes. In a full derivation of canonical effective
equations of quantum gravity or one of its models, moment equations
accompany those for expectation values; it is not necessary to perform
any additional quantization to derive such equations. In this way,
initial conditions for inflationary scenarios can be derived at the
effective level, even though the ground or vacuum state is not a
semi-classical one. The resulting vacuum choice is not unique, since
in the presence of loop quantum corrections different prescriptions
can give rise to the same classical limit
\cite{LoopMuk,HolState}. This is not a matter of particular concern,
as the uniqueness of the cosmological vacuum is a prerogative of
standard general relativity and, in general scenarios where the
dispersion relation or other ingredients are modified, one must make a
specific choice based on some reasonable assumptions.

This derivation has an additional advantage, not just the one of avoiding the
use of wave functions. In order to find the conditions of a Bunch--Davies
vacuum, one must use quantum field theory in curved space-time. In deformed
general relativity, for which loop quantum gravity provides examples,
space-time does not have the classical structure, and it is not known how to
perform standard quantum field theory on it (see \cite{Energy}). The canonical
effective equations used here, on the other hand, are still available, so that
the correct initial conditions for vacuum fluctuations can be derived from
moments. In particular, the deformation function $\beta$, or rather
$\sqrt{|\beta|}$ then appears in the normalization via $\Delta v_k$,
multiplying $k$. Instead of (\ref{MS}), one would have $v_k''+(\beta
    k^2-z''/z)v_k=0$ from (\ref{EoM}). This correction of the
normalization condition agrees with the WKB calculations of \cite{HolState}.

\subsection{Quantum-gravity state}

In models of quantum gravity, states must be selected or derived not just for
matter or the inflaton but also for gravitational degrees of
freedom. Unfortunately, distinguished classes of such states are much more
difficult to find than for matter theories because the gravitational
Hamiltonian is unbounded from below and does not allow a ground state. And
even if a ground state existed under some additional conditions, most likely
it would not provide a good choice of state in high-density regimes near the
big bang. Gaussian states have often been considered in homogeneous cosmology
and shown to capture the main semi-classical features of loop models provided
the state remains sharply peaked at high density. In the solvable model of
flat, isotropic space with a free, massless scalar, a sharply peaked state at
large volume remains sharply peaked at high density, and semiclassical
features are sufficient to conclude that there is a bounce. This conclusion,
restricted to the solvable model, even holds for states more general than
Gaussians \cite{BouncePert,DGMS}. However, for non-solvable models and
especially inhomogeneous space-times the choice of state is more
delicate. Using a simple Gaussian or some other form of coherent or
semi-classical wave function is not fully justified, in contrast with the case
of the initial inflaton state.

If one does assume a Gaussian to evolve through a possible bounce, one
presupposes that the state remains nearly classical even in a strong quantum
regime where, for instance, holonomy modifications should be
important. Therefore, one implicitly assumes that there is no strong quantum
back-reaction, or no higher-curvature terms in addition to those higher-order
modifications put in by holonomies. It is then not surprising that such states
stay near those of simple holonomy-modified classical equations such as
(\ref{FriedHol}), which show a bounce by the bound imposed on the energy
density. However, quantum back-reaction is ignored by choosing a Gaussian, and
therefore it is not guaranteed that the bounces shown by (\ref{FriedHol}) are
reliable in loop quantum cosmology, even setting aside the issue of signature
change. 

A full understanding of the Planck regime in loop quantum gravity requires a
detailed analysis of the dynamics of generic states. Such investigations are
in principle possible using wave functions, but they have not been started
yet. As long as these results remain lacking, one cannot be sure of the
quantum nature of the big bang, not just about details but even about general
questions such as whether there is a bounce. An analysis of generic states and
their approach to the Planck regime would be easier to perform using the
moment behavior described by canonical effective equations because the number
of state parameters to be controlled would be smaller at any fixed order in
the semi-classical expansion. Also this task, however, still remains to be
completed. For now, therefore, the main cosmological lesson to be drawn from
loop quantum gravity is the possible form of deformed space-time structures
and their implications, but not any detailed scenario such as a bounce.

\section{Conclusions}

We have presented a detailed discussion of effective methods for models of
canonical quantum gravity in relation to the anomaly problem. We have mainly
discussed the overall coherence and consistency of the framework, but also
provided new details and insights: (i) We have compared different anomaly-free
models and showed their agreements, (ii) have considered higher-order
perturbations in this framework, and (iii) have described an effective way of
specifying an initial inflaton state.

Our results pave the way to several new developments, all of which
  would require more time and space than is available for this
  article. However, while important technical details still have to be
  completed in these directions, the conceptual discussions presented here have
  removed what had appeared to be difficult hurdles. In particular, our
  description of a framework for anomaly-free higher-order perturbations will
  be important for attempts to derive non-Gaussianity from models of loop
  quantum cosmology. Specific calculations would require long manipulations of
  Poisson brackets of parameterized constraints with, say, holonomy
  modifications. If anomaly-free constraints existed to higher orders, their
  derivation would follow the scheme provided here. If such constraints did not
  exist, the same scheme would allow one to test whether new kinds of
  corrections could provide anomaly-freedom. Our characterization of an
  effective description of initial inflaton states should be easier to
  implement. Here, the main new direction would be an exploration of effects
  from non-vacuum states. More generally, the construction of an anomaly-free perturbation dynamics will permit the extraction of the full set inflationary observables in loop quantum cosmology, thus completing extant results based on a partial or preliminary implementation of quantum corrections.

As emphasized, although a general framework to derive effective theories of
canonical quantum gravity is available, it remains incompletely realized in
present concrete calculations. Especially higher spatial and higher time
derivatives must still be included in the existing models. No alternative
procedure has yet yielded such terms for models of loop quantum gravity, and
therefore it is unclear if they can be consistent at all. (If they cannot,
there would be a significant mismatch between loop quantum gravity and general
expectations from effective field theory, as realized in perturbative quantum
gravity \cite{EffectiveGR,BurgessLivRev}.) Moreover, several features we
described are lacking in such alternative approaches. We propose the effective
approach sketched here as a candidate complete method to address
the anomaly problem and to extract physical predictions from canonical quantum
gravity.

\section*{Acknowledgements}

We thank Ivan Agull\'o, Jaume Garriga, Guillermo Mena-Marug\'an and Edward
Wilson-Ewing for discussions.  This work was supported in part by NSF grants
PHY-0748336 and PHY-1307408. The work of G.C.\ is under a Ram\'on y Cajal
contract.

\begin{appendix}

\section{Canonical effective theory}
\label{App:Moments}

A canonical formulation of effective equations can be obtained by working with
expectation values and moments of states with respect to a basic set of
operators.  One can introduce a Poisson bracket on the space of expectation
values by
\begin{equation}\label{Poisson}
\{\langle\hat{A}\rangle,\langle\hat{B}\rangle\}:=
\frac{\langle[\hat{A},\hat{B}]\rangle}{i\hbar}\,.
\end{equation}
For basic operators, this Poisson bracket agrees with the classical one, by
virtue of the standard quantization procedure. Imposing the Leibniz rule, the
Poisson bracket is extended to fluctuations and moments of basic operators,
providing non-classical degrees of freedom. If moments at arbitrary order are
defined by 
\begin{equation}
 \Delta(q^ap^b):= \left\langle \left[ (\hat{q}-\langle\hat{q}\rangle)^a
     (\hat{p}-\langle\hat{p}\rangle)^b\right]_{\rm symm}\right\rangle
\end{equation}
with totally symmetric ordering, they all Poisson commute with expectation
values $\langle\hat{q}\rangle$ and $\langle\hat{p}\rangle$ of basic operators,
provided the latter form a canonical algebra with $[\hat{q},\hat{p}]$ a
constant. The moments are therefore truly independent phase-space degrees of
freedom. (For a non-canonical algebra of basic variables, which sometimes
appears in loop models when holonomies are not expanded, the phase-space
structure is more complicated. However, at the perturbative level, an
expansion of holonomies is required because higher powers of the connection in
such an expansion will be mixed with moment terms contributing to higher time
derivatives. The assumption of a canonical basic algebra is therefore a safe
one.) For instance, two fluctuations have a Poisson bracket
\begin{eqnarray*}
 \{(\Delta A)^2,(\Delta B)^2\} &=&
 \{\langle\hat{A^2}\rangle-\langle\hat{A}\rangle^2,
 \langle\hat{B}^2\rangle-\langle\hat{B}\rangle^2\}\\
&=& \frac{1}{i\hbar} \left( \langle[\hat{A}^2,\hat{B}^2]\rangle-
  2\langle\hat{A}\rangle \langle[\hat{A},\hat{B}^2]\rangle-
  2\langle\hat{B}\rangle\langle[\hat{A}^2,\hat{B}]\rangle+
  4\langle\hat{A}\rangle\langle\hat{B}\rangle
\langle[\hat{A},\hat{B}]\rangle\right)\\
&=& \{\langle\hat{A}^2\rangle,\langle\hat{B}^2\rangle\}-
2\langle\hat{A}\rangle \{\langle\hat{A}\rangle,\langle\hat{B^2}\rangle\}-
2\langle\hat{B}\rangle \{\langle\hat{A}^2\rangle,\langle\hat{B}\rangle\}\nonumber\\
&&+4\langle\hat{A}\rangle \langle\hat{B}\rangle
\{\langle\hat{A}\rangle,\langle\hat{B}\rangle\}\,,
\end{eqnarray*}
fully expressed in terms of (\ref{Poisson}).

The Heisenberg equation for an operator $\hat{O}$ is equivalent to a
Hamiltonian equation for the expectation value $\langle\hat{O}\rangle$ with
effective Hamiltonian $\langle\hat{H}\rangle$:
\begin{equation} \label{dOdt}
\frac{{\rm d}\langle\hat{O}\rangle}{{\rm d}t} =
\frac{\langle[\hat{O},\hat{H}]\rangle}{i\hbar}=
\{\langle\hat{O}\rangle,\langle\hat{H}\rangle\}\,.
\end{equation}
In order to write $\langle\hat{H}\rangle$ explicitly as a function of
expectation values of basic operators and their moments, one may expand as
\begin{eqnarray} \label{HQ}
\langle\hat{H}\rangle &=& \langle H[\langle\hat{q}\rangle+
(\hat{q}-\langle\hat{q}\rangle), \langle\hat{p}\rangle+
(\hat{p}-\langle\hat{p}\rangle)]\rangle\\
&=& H(\langle\hat{q}\rangle,\langle\hat{p}\rangle)+ \sum_{a+b=2}^{\infty}
 \frac{1}{a!b!}
 \frac{\partial^{a+b}H(\langle\hat{q}\rangle, \langle\hat{p}\rangle)}{\partial
  \langle\hat{q}\rangle^a \partial \langle\hat{p}\rangle^b} \Delta(q^ap^b)\,,
\nonumber
\end{eqnarray}
assuming that the Hamiltonian operator $\hat{H}$ is ordered totally
symmetrically. If it is not, one can relate it to a totally symmetric
operator by adding terms to (\ref{HQ}) that explicitly depend on $\hbar$
\cite{Casimir,Search}.

These effective methods apply to constrained systems by using
$\langle\hat{C}\rangle$ (as well as $\langle\widehat{{\rm pol}}\hat{C}\rangle$
for operators $\widehat{{\rm pol}}$ polynomial in
$\hat{q}-\langle\hat{q}\rangle$ and $\hat{p}-\langle\hat{p}\rangle$) instead
of $\langle\hat{H}\rangle$ \cite{EffCons,EffConsRel}.  An extension to quantum
field theory is in progress, but many questions, including those studied in
this article, can be addressed at a regularized level (such as states with a
fixed graph in loop quantum gravity) with methods developed for a finite
number of degrees of freedom.

\section{Consistency of inverse-triad corrected equations}
\label{App:Consist}

To illustrate the idea that the consistency of quantum-corrected equations
cannot be taken for granted and needs to be checked, we consider the following
example. In \cite{ConstraintAlgebra} the cosmological equations of motion
including inverse-triad corrections have been derived. Below we explicitly
demonstrate that the set of background and perturbed equations is
consistent. To lighten the notation, we removed all bars from background
functions.

The quantum corrected background equations of motion are given by
\begin{eqnarray}
{\cal H}^2&=&\frac{8\pi G}{3}\alpha\left[\frac{\dot{
\varphi}^2}{2\nu}+{p}V(\varphi)\right]\label{FriBG},\\
\dot{\cal H}&=&{\cal H}^2\left(1+\frac{\alpha^\prime 
p}{\alpha}\right)-4\pi
G\frac{\alpha}{\nu}\dot{\varphi}^2\left(1-\frac{\nu^\prime
 p}{3\nu}\right)\label{RayBG},\\
\ddot{\varphi}&+&2{\cal H}\dot{\varphi}\left(1-\frac{\nu^\prime
p}{\nu} \right)+\nu  p
V_{,\varphi}(\varphi)=0\label{KGBG}\,,
\end{eqnarray}
where, only in this appendix, the dot denotes a conformal time
derivative and the prime indicates a derivative with respect to $
p=a^2$.  Classically, only two of the three equations are
independent. Here we demonstrate that it is still the case with
quantum corrections.

We start by recalling a convenient auxiliary relation for an
arbitrary function $F$ of $ p$
\begin{equation}\label{dot_prime}
[F( p)]^{\bullet}=2{\cal H}  p F^\prime( p)\,.
\end{equation}
Now take a conformal time derivative of the Friedmann equation
(\ref{FriBG}). The resulting $\dot{\cal H}$ and $V(\varphi)$-terms can be
eliminated using Eqs.~(\ref{RayBG}) and (\ref{FriBG}), respectively. It is
then easy to see that bringing all the terms on one side of the equation and
dividing them by $(8\pi G/3)(\alpha/\nu)\dot\varphi$ yields the background
Klein--Gordon equation (\ref{KGBG}).

Given anomaly-freedom of the constraints, it is possible to construct
gauge-invariant variables and recast the equations of motion in a manifestly
gauge-invariant manner. This analysis has been done for the perturbative
constraints of \cite{ConstraintAlgebra}, incorporating inverse-triad
corrections of loop quantum gravity in a way which, to leading order, is
anomaly-free. The final equations in terms of the scalar perturbations $\Phi$ and $\Psi$,
\begin{eqnarray}
&&\partial_c\left[\dot\Psi+{\cal H}(1+f)\Phi\right]=4\pi
G\frac{{\alpha}}{{\nu}}\dot\varphi \partial_c\delta\varphi\,,\label{Discussion_Diff}\\
&&\nabla^2({\alpha}^2
\Psi)-3{\cal H}(1+f)\left[\dot\Psi+{\cal H}(1+f)\Phi\right]=4\pi
G\frac{{\alpha}}{{\nu}}(1+f_3)\left[\dot{{\varphi}}
\delta\dot\varphi-\dot{{\varphi}}^2(1+f_1)\Phi\right.\nonumber\\
&&\qquad\qquad \left.+{\nu} {p} V_{,\varphi}({\varphi})\delta\varphi\right]\,,\label{Discussion_Ham}\\
&&4\pi G\frac{{\alpha}}{{\nu}}\left[\dot\varphi\delta\dot\varphi-{p}{\nu} V_{,\varphi}({\varphi})\delta\varphi\right]
= \ddot\Psi+{\cal H}\left[2\dot\Psi\left(1-\frac{{\alpha}^\prime
{p}}{{\alpha}}\right)+\dot\Phi(1+f)\right]\nonumber\\
&&\qquad\qquad+\left[\dot{\cal H}+2{\cal H}^2\left(1+f^\prime {p} -
\frac{{\alpha}^\prime {p}}{{\alpha}}\right)\right]\Phi(1+f),\nonumber\\\label{Discussion_Ray}\\
&&\partial_a\partial^i\left[{\alpha}^2\left(\Phi-\Psi(1+h)\right)\right]=0\,,\label{Discussion_OffDiag}\\
&&\delta \ddot \varphi+2 {\cal H} \delta \dot \varphi \left(1 -
\frac{{\nu}^\prime {p}}{{\nu}}-g_1^\prime {p}\right)-
{\nu} {\sigma} (1-f_3)\nabla^2 \delta \varphi +{\nu}
{p} V,_{\varphi\varphi}({\varphi})\delta \varphi\label{Disscussion_KG}\\
&&\qquad\qquad+2\left[{\nu} {p}
V,_{\varphi}({\varphi})(1+f_1)-{\cal H}\dot{{\varphi}}
(f_3^\prime {p})\right] \Phi -
\dot{{\varphi}}\left[(1+f_1)\dot\Phi
+3(1+g_1)\dot\Psi\right]=0\,,\nonumber
\end{eqnarray}
with specific relationships between the correction functions $f$, $f_i$,
  $g_1$, $h$, $\alpha$ and $\nu$ (see below), are manifestly gauge invariant and reproduce the classical perturbed Einstein
equation if one omits the quantum corrections. If the constraints had not been
ensured to be anomaly-free, the resulting perturbation equations would couple
gauge-dependent terms to the supposedly gauge-invariant curvature
perturbation. Apart from gauge invariance, there is also a consistency issue
which arises since, on general grounds, there are three unknown scalar
functions subject to five equations. Moreover, with suitable spatial boundary
conditions, Eqs.~(\ref{Discussion_Diff}) and (\ref{Discussion_OffDiag}) can be
used to eliminate two of these functions in terms of just one, say $\Psi$,
which should satisfy the three remaining equations.

Another consistency check that is available classically is comparing the
diffeomorphism constraint equation (\ref{Discussion_Diff}) and the perturbed
Raychaudhuri equation (\ref{Discussion_Ray}). Specifically, one can remove the
gradient from Eq.~(\ref{Discussion_Diff}), again assuming suitable spatial
boundary conditions, and take a conformal time derivative. Classically, the
resulting equation is equivalent to Eq.~(\ref{Discussion_Ray}). Importantly,
this consistency remains true in the presence of quantum corrections:
Eq.~(\ref{Discussion_Diff}) implies
\begin{equation}\label{Discussion_Diff_no_gradient}
\dot\Psi+{\cal H}(1+f)\Phi=4\pi G\frac{{\alpha}}{{\nu}}\dot\varphi\,
\delta\varphi\,,
\end{equation}
whose conformal time derivative yields
\[
\ddot\Psi+\dot{\cal H}(1+f)\Phi+{\cal H}\dot f\Phi+{\cal H}(1+f)\dot\Phi=4\pi
G\left(\frac{{\alpha}}{{\nu}}\dot\varphi
\delta\varphi\right)^{\bullet}\,.
\]
Subtracting Eq.~(\ref{Discussion_Ray}) and replacing the time derivative $\dot
f$ with $f^\prime \dot{ p} \equiv 2{\cal H} f^\prime  p$, we
obtain
\begin{equation}
 \label{Diff_Ray}-2{\cal H}^2\Phi\left(1+f-\frac{\alpha^\prime 
p}{\alpha}\right)-2{\cal H}\dot\Psi\left(1-\frac{\alpha^\prime 
p}{\alpha}\right)=4\pi G
\left[\left(\frac{{\alpha}}{{\nu}}\dot\varphi
\delta\varphi\right)^{\dot{}}-\frac{\alpha}{\nu}\left(\dot{\varphi}\delta\dot\varphi-{p}{\nu}
V_{,\varphi}({\varphi})\delta\varphi\right)\right].
\end{equation}
Using Eq.~(\ref{Discussion_Diff_no_gradient}), the left-hand side can
be rewritten as
\begin{eqnarray}
-2{\cal H}^2\Phi\left(1+f-\frac{\alpha^\prime 
p}{\alpha}\right)-2{\cal H}\dot\Psi\left(1-\frac{\alpha^\prime 
p}{\alpha}\right)&=&-2{\cal H}\left(1-\frac{\alpha^\prime 
p}{\alpha}\right)\left[\dot\Psi+{\cal H}\Phi(1+f)\right]\nonumber\\
&=&-2{\cal H}\left(1-\frac{\alpha^\prime  p}{\alpha}\right)4\pi
G\frac{{\alpha}}{{\nu}}\dot{\varphi} \delta\varphi\,.
\nonumber
\end{eqnarray}
After cancelling the factor of $4\pi G$, Eq.~(\ref{Diff_Ray})
becomes
\[
-2{\cal H}\left(1-\frac{\alpha^\prime
    p}{\alpha}\right)\frac{{\alpha}}{{\nu}}\dot{\varphi}
\delta\varphi=\left(\frac{{\alpha}}{{\nu}}\dot{\varphi}
  \delta\varphi\right)^{\bullet}-\frac{\alpha}{\nu}\left(\dot{\varphi}\delta\dot\varphi-{p}{\nu}
  V_{,\varphi}({\varphi})\delta\varphi\right)\,.
\]
It is easy to see that the terms containing the time derivative of
the matter perturbation mutually cancel. Bringing all the
remaining terms to one side of the equation results in
\begin{equation}\label{Diff_Ray_1}
\delta\varphi\left[\left(\frac{ \alpha}{
\nu}\dot{\varphi}\right)^{\bullet}+2{\mathscr
H}\frac{\alpha}{\nu}\left( 1-\frac{\alpha^\prime
p}{\alpha}\right)\dot{\varphi}+\alpha p
V_{,\varphi}(\varphi)\right]=0\,.
\end{equation}
Since $( \alpha/ \nu)^{\bullet}\equiv 2( \alpha/ \nu)'
{\cal H} p=2{\cal H}(\alpha/\nu)(\alpha^\prime
p/\alpha-\nu^\prime  p/\nu)$, the expression inside the square
brackets
\[
 \frac{\alpha}{\nu}\left[\ddot{\varphi}+2{\mathscr
     H}\dot{\varphi}\left(1-\frac{\nu^\prime { p}}{\nu} 
\right)+\nu  p V_{,\varphi}(\varphi)\right]
\] 
is exactly the quantum corrected background Klein-Gordon equation
(\ref{KGBG}). Therefore Eq.~(\ref{Diff_Ray_1}), which constitutes the
difference between the time derivative of (\ref{Discussion_Diff_no_gradient})
and Eq.~(\ref{Discussion_Ray}), is identically satisfied, indicating
equivalence of the diffeomorphism equation and the perturbed Raychaudhuri
equation.

One more perspective on closure of the equations of motion is given by showing
that the Klein--Gordon equation (\ref{Disscussion_KG}) is not independent. In
the covariant formalism, it results from the energy conservation equation for
the matter field: $\nabla_\mu T^{\mu\nu}=0$, the counterpart of the Bianchi
identity of the gravitational sector. The latter equation is automatically
satisfied by construction of the Einstein tensor. For this reason, the
Klein--Gordon equation can be expressed in terms of the other equations and
their derivatives.  In the canonical formulation such an argument, referring
to the Bianchi identity, is unfortunately not available, especially at the
effective level, for it is a priori not clear what kind of action or
space-time structure might correspond to the quantum corrected constraints.
Nonetheless, we shall demonstrate the redundancy of the Klein--Gordon equation
by an explicit derivation below.

As we will make an extensive use of the anomaly freedom conditions
presented in \cite{ConstraintAlgebra} and Appendix B of
\cite{ScalarGaugeInv}, it is convenient to summarize them here:
\begin{eqnarray}
\label{Anom_B2} h+f-2\frac{\alpha^\prime  p}{\alpha}&=&0\,,\\
\label{Anom_B3} 2f^\prime p+\frac{\alpha^\prime 
p}{\alpha}&=&0\,,\\
\label{Anom_B5} g_1+f_3-f_1&=&0\,,\\
\label{Anom_B6} f-f_1-\frac{\nu^\prime  p}{3\nu}&=&0\,,\\
\label{Anom_B8} 2f_3^\prime p+ 3(f_3-f)&=&0\,,\\
\label{Anom_B1_phi} (f_1+f_3)^\prime  p+ \frac{\nu^\prime 
p}{\nu}&=&0\,.
\end{eqnarray}
We start by noting that
\[
(1+f)\Phi = (1+f+h)\Psi = \left(1+2\frac{\alpha^\prime
p}{\alpha}\right)\Psi = \frac{2\sqrt{
p}}{\alpha}(\alpha\sqrt{ p})^{\prime}\Psi = \frac{(\alpha
\sqrt{ p})^{\bullet}}{\alpha\sqrt{ p}{\cal H}}\Psi\,,
\]
where we have used (\ref{Discussion_OffDiag}) to replace $\Phi$
with $\Psi$ and (\ref{Anom_B2}). Using this result,
Eq.~(\ref{Discussion_Diff_no_gradient}) can be rewritten in the
compact form
\begin{equation}\label{Diff_compact}
\left(\alpha\sqrt{ p}\Psi\right)^{\bullet}=4\pi G
\frac{\alpha^2}{\nu}\sqrt{ p}\dot\varphi\delta\varphi\,.
\end{equation}
Also, using (\ref{Discussion_Diff_no_gradient}) to eliminate the
second term in Eq.~(\ref{Discussion_Ham}), the latter can be
rewritten as
\begin{equation}\label{Ham_compact}
\nabla^2\left(\alpha\sqrt{ p}\Psi\right)=4\pi G \frac{\sqrt{
p}}{\nu}(1+f_3)
\left[\dot\varphi\delta\dot\varphi-\dot\varphi^2(1+f_1)\Phi+(\nu
 p V_{,\varphi}+3{\cal H}\dot\varphi(1+f-f_3))\delta\varphi\right]\,.
\end{equation}
Eqs.~(\ref{Diff_compact}) and (\ref{Ham_compact}) are analogous to
the perturbed diffeomorphism and Hamiltonian constraint equations
respectively. The difference
$\left[(\ref{Ham_compact})^{\bullet}-\nabla^2(\ref{Diff_compact})\right]$
is equivalent to the perturbed Klein-Gordon equation. Indeed,
consider
\begin{eqnarray}
 0&=&\frac{1}{4\pi G}\left[(\ref{Ham_compact})^{\bullet}-\nabla^2(\ref{Diff_compact})\right]\nonumber\\
&=&\left\{\frac{\sqrt{p}}{\nu}(1+f_3)\left[\dot\varphi\delta\dot\varphi-\dot\varphi^2(1+f_1)\Phi+(\nu  p
V_{,\varphi}+3{\cal H}\dot\varphi(1+f-f_3))\delta\varphi\right]\right\}^{\bullet}\nonumber\\
&&-\frac{\alpha^2}{\nu}\sqrt{ p}\dot\varphi\nabla^2\delta\varphi\label{Equiv_KG}
\end{eqnarray}
and collect similar terms on the right-hand side:
\begin{description}
\item[$\delta \ddot\varphi$-terms:]
\begin{equation}
 \delta\ddot\varphi\left\{\dot\varphi(1+f_3)\frac{\sqrt{
 p}}{\nu}\right\}\label{phi_ddot_term}\,.
\end{equation}
\item[$\nabla^2\delta \varphi$-terms:]%
\begin{equation}
 \nabla^2\delta \varphi\left\{-\dot\varphi\frac{\alpha^2\sqrt{
 p}}{\nu}\right\}\label{lapl_phi_term}\,.
\end{equation}
\item[$\delta \dot\varphi$-terms:]%
\begin{eqnarray}
&& \delta \dot\varphi\left\{\left[\dot\varphi(1+f_3)\frac{\sqrt{
p}}{\nu}\right]^{\bullet}+
 \frac{\sqrt{ p}}{\nu}(1+f_3)\left[\nu  p
 V_{,\varphi}+3{\cal H}\dot\varphi(1+f-f_3)\right]\right\}\nonumber\\
&& =\delta \dot\varphi\left\{2{\cal H}\dot\varphi\left(1+f_3+2f_3^\prime
 p\right)\right\}\,,\label{phi_dot_term}
\end{eqnarray}
where we have used the background Klein--Gordon equation (\ref{KGBG}) to get
rid of the $\ddot\varphi$-term and the anomaly cancellation condition
(\ref{Anom_B8}) to simplify the final expression.
\item[$\delta \varphi$-terms:]%
\begin{eqnarray}
&&\delta \varphi\left\{\frac{\sqrt{ p}}{\nu}(1+f_3) \left[\nu
 p
V_{,\varphi}+3{\cal H}\dot\varphi(1+f-f_3)\delta\varphi\right]\right\}^{\bullet}\nonumber\\
&&=\delta \varphi\left\{\left[
p^{3/2}(1+f_3)V_{,\varphi}\right]^{\bullet}+\left[3{\cal
  H}\dot\varphi(1+f)\frac{\sqrt{p}}{\nu}\right]^{\bullet}\right\}\nonumber\\
&&=\delta \varphi\Biggl\{ p^{3/2}(1+f_3)
V_{,\varphi\varphi}\dot\varphi+{\cal H}
p^{3/2}V_{,\varphi}\left[3(1+f_3)+2f_3^\prime\right]+3{\cal H}\ddot\varphi(1+f)\frac{\sqrt{
p}}{\nu}\nonumber\\
&&\quad\quad\quad+3\dot{\cal H}\dot\varphi(1+f)\frac{\sqrt{
p}}{\nu}+3{\cal H}^2\dot\varphi\frac{\sqrt{
p}}{\nu}\left[1-\frac{2\nu^\prime p}{\nu}+f+2f^\prime
p\right]\Biggr\}\nonumber\\
&&=\delta \varphi\left\{ p^{3/2}(1+f_3)
V_{,\varphi\varphi}\dot\varphi-3\frac{\dot\varphi^2}{\delta\varphi}\frac{\sqrt{
p}}{\nu}(1+f_1)\left[\dot\Psi+{\mathscr
H}(1+f)\Phi\right]\right\}\,,\label{del_phi_term} 
\end{eqnarray}
where we again eliminated the $\ddot\varphi$-term using
(\ref{KGBG}), substituted $\dot{\cal H}$ from the background
Raychaudhuri equation (\ref{RayBG}) in which the $(4\pi
G\dot\varphi)$-factor was expressed using
(\ref{Discussion_Diff_no_gradient}). We also applied the anomaly
freedom conditions (\ref{Anom_B3}), (\ref{Anom_B6}) and
(\ref{Anom_B8}) to obtain the last line. Note that the second term
contributes to the $\Phi$- and $\dot \Psi$-terms below.
\item[$\Phi$-terms:]%
\begin{eqnarray}
&&\Phi\left\{-\left[\frac{\sqrt{
p}}{\nu}(1+f_1+f_3)\dot\varphi^2\right]^{\bullet}-3{\cal
H}\dot\varphi^2\frac{\sqrt{p}}{\nu}(1+f+f_1)\right\}\nonumber\\
&&=\Phi\left\{2
p^{3/2}\dot\varphi(1+f_1+f_3)-2{\cal H}\dot\varphi\frac{\sqrt{
p}}{\nu}(f_3^\prime p)\right\}\,,\label{Phi_term}
\end{eqnarray}
where as before we have used (\ref{KGBG}) to substitute for
$\ddot\varphi$ and (\ref{Anom_B8}) to simplify the final
expression. Note also that the second term in the first line
originates from Eq.~(\ref{del_phi_term}).
\item[$\dot\Phi$-terms:]%
\begin{eqnarray}\label{Phi_dot_term}
\dot\Phi\left\{-\frac{\sqrt{
p}}{\nu}(1+f_1+f_3)\dot\varphi^2\right\}\,.
\end{eqnarray}
\item[$\dot\Psi$-terms:]%
\begin{eqnarray}\label{Psi_dot_term}
\dot\Psi\left\{-3\frac{\sqrt{
p}}{\nu}(1+f_1)\dot\varphi^2\right\}=\dot\Psi\left\{-3\frac{\sqrt{
p}}{\nu}(1+f_3+g_1)\dot\varphi^2\right\}\,.
\end{eqnarray}
\end{description}
This contribution comes entirely from Eq.~(\ref{del_phi_term}) and the
equality holds by virtue of the anomaly freedom condition
(\ref{Anom_B5}).
Dividing each of (\ref{phi_ddot_term})--(\ref{Psi_dot_term}) by
$(\sqrt{ p}/\nu)(1+f_3)\dot\varphi$ and neglecting
subleading orders of Planck's constant, we see that
(\ref{Equiv_KG}) reads
\begin{eqnarray}
\delta \ddot \varphi &+&2 {\cal H} \delta \dot \varphi \left(1 +
2f_3^\prime p\right) + {\nu} {\sigma} (1-f_3)\nabla^2
\delta \varphi +{\nu} {p}
V,_{\varphi\varphi}({\varphi})\delta \varphi
\label{Final_KG}\\
&+&2\Phi\left[{\nu} {p} V,_{\varphi}({\varphi})(1+f_1)
-{\cal H}\dot{{\varphi}} (f_3^\prime {p})\right]
-\dot{{\varphi}}\left[(1+f_1)\dot\Phi
+3(1+g_1)\dot\Psi\right]=0\,,\nonumber
\end{eqnarray}
which is equivalent to the perturbed Klein--Gordon equation
(\ref{Disscussion_KG}), since $\nu\sigma = \alpha^2$, and the
coefficient in front of the $\delta \dot \varphi$-term in
(\ref{Final_KG}) can be rewritten, by virtue of (\ref{Anom_B2})--(\ref{Anom_B1_phi}), in several equivalent forms:
\[
1+2f_3^\prime p= 1-\frac{\nu^\prime  p}{\nu}-g_1^\prime 
p = 1+3(f-f_3)=1+\frac{\nu^\prime  p}{\nu}+3g_1\,.
\]
Note that the second expression coincides with the coefficient in
(\ref{Disscussion_KG}).

It is also possible to derive an equation not containing matter
fields  by subtracting (\ref{Discussion_Ham}) divided by $(1+f_3)$
from (\ref{Discussion_Ray}). Eliminating $V_{,\varphi}$,
$\dot\varphi^2$ and $\delta\varphi$ using the background
Klein--Gordon equation (\ref{KGBG}), background Raychaudhuri
equation (\ref{RayBG}) and the perturbed diffeomorphism equation
(\ref{Discussion_Diff_no_gradient}) respectively one obtains \
\begin{eqnarray}\label{Mukhanov}
\ddot\Psi&-&\alpha^2(1-f_3)\nabla^2\Psi+2\dot\Psi\left[{\cal H}\left(1+f_3^\prime
p+2\frac{\nu^\prime
p}{\nu}\right)-\frac{\ddot\varphi}{\dot\varphi}\right]\\
&+&2\Psi\left[\dot{\cal H}\left(1+\frac{2\alpha^\prime}{\alpha}\right)+{\cal H}^2\left(2\frac{\alpha^{\prime\prime}
p^2}{\alpha}+2\frac{\nu^\prime p}{\nu}+(f_3-f)^\prime
p\right)-{\cal H}\frac{\ddot\varphi}{\dot\varphi}\left(1+\frac{2\alpha^\prime
 p}{\alpha}\right)\right]=0\,.\nonumber
\end{eqnarray}
In order to arrive at the last expression, the anomaly
cancellation conditions (\ref{Anom_B2}), (\ref{Anom_B3}),
(\ref{Anom_B6}), and (\ref{Anom_B8}) have been used, along with
the fact that
\[
\left(\frac{\alpha^\prime 
p}{\alpha}\right)^\prime  p=\frac{\alpha^{\prime\prime} 
p^2}{\alpha}+\frac{\alpha^\prime  p}{\alpha}
\]
up to higher order terms in Planck's constant. Interestingly,
Eq.~(\ref{Mukhanov}) can also be obtained by replacing all the
$\delta\varphi$- and $\delta\dot\varphi$-terms in the perturbed
Hamiltonian constraint equation (\ref{Discussion_Ham}) by the
corresponding expressions according to
(\ref{Discussion_Diff_no_gradient}). Consistency of the original
equations demonstrated  above ensures that one will arrive at the
same equation (\ref{Mukhanov}).

As these explicit calculations demonstrate, consistency crucially relies
  on tight relationships between different corrections in the mode
  equations. Moreover, corrections in mode equations must be closely related
  to corrections in the background equations. These relationships follow from
  the condition of anomaly-freedom as realized by a closed constraint
  algebra. If one were to fix the gauge before quantization or to modify the
  equations for classical gauge-invariant variables, these tight relations
  would be overlooked and corrections would be much more ambiguous.

\end{appendix}


\end{document}